\documentclass[preprintnumbers,nofootinbib,eqsecnum,aps,prd]{revtex4}

\usepackage{graphicx,epstopdf}
\usepackage{amsmath,amssymb,amsfonts}
\usepackage{dcolumn}
\usepackage{bm}

\newcommand{\md}{\mathrm{d}}
\newcommand{\mpl}{M_{P}}
\newcommand{\ba}{\begin{eqnarray}}
\newcommand{\ea}{\end{eqnarray}}
\newcommand{\bmk}{{\bm k}}

\newcommand{\bmp}{{\bm p}}

\newcommand{\bmx}{{\bm x}}
\newcommand{\calN}{{\cal N}}
\newcommand{\calR}{{\cal R}}

\begin{document}

\title{Curvature perturbation and waterfall dynamics in hybrid inflation}
\author{Ali Akbar Abolhasani$^{1,3}$}
\email{abolhasani-AT-mail.ipm.ir}
\author{Hassan Firouzjahi$^{2}$}
\email{firouz-AT-mail.ipm.ir}
\author{Misao Sasaki$^{3,4}$}
\email{misao-AT-yukawa.kyoto-u.ac.jp}
\affiliation{$^1$Department of Physics, Sharif University of 
Technology, Tehran, Iran}
\affiliation{$^2$School of Physics, Institute for Research in 
Fundamental Sciences (IPM),
P. O. Box 19395-5531,
Tehran, Iran}
\affiliation{$^3$ Yukawa Institute for theoretical Physics,
 Kyoto University, Kyoto 606-8502, Japan}
\affiliation{$^4$ Korea Institute for Advanced Study,
207-43 Cheongnyangni 2-dong, Dongdaemun-gu,
Seoul 130-722, Republic of Korea}
\date{\today}

\begin{abstract}

We investigate the parameter spaces of hybrid inflation model with special 
attention paid to the dynamics of waterfall field and curvature perturbations
induced from its quantum fluctuations.  Depending on the inflaton field value
at the time of phase transition and the sharpness of the phase transition
inflation can have multiple extended stages. We find that for models with
mild phase transition the induced curvature perturbation from the waterfall
field is too large to satisfy the COBE normalization.
We investigate the model parameter space where the curvature perturbations
from the waterfall quantum fluctuations vary between the results of standard
hybrid inflation and the results obtained here.

\end{abstract}

\preprint{YITP-11-4, IPM/P-2011/003}

\maketitle
\section{Introduction}

Thanks to vast data obtained from recent cosmological observations such as 
WMAP7 \cite{Komatsu:2010fb} inflation \cite{Guth:1980zm} has emerged as the
 leading theory of the early universe and structure formation.
 The simplest class of models of inflation is based on a scalar field,
the inflaton field, coupled minimally to gravity~\cite{Linde:1981mu,Linde:1983gd}. 
The inflaton field potential is flat enough to support long enough period 
of inflation to solve the horizon and flatness problems associated 
with the big bang cosmology.
These simple models predict almost scale invariant, almost Gaussian and
 almost adiabatic perturbations
 which are very well consistent with the
 Cosmic Microwave Background (CMB) observations~\cite{Komatsu:2010fb}.
However, with the help of 
ever increasing accurate cosmic measurements different inflationary models
 can be discriminated based on their predictions for the power spectrum 
spectral index or the amount of gravitational waves.


Recently there has been a revival of interest in hybrid
inflation~\cite{Linde:1993cn,Copeland:1994vg}, specifically the dynamics 
of waterfall field and whether or not significant large scale curvature 
perturbations can be produced during the waterfall phase 
transition~\cite{Lyth:2010ch, Abolhasani:2010kr, Fonseca:2010nk,
 Gong:2010zf, Lyth:2010zq, Levasseur:2010rk}, for earlier works see also
 \cite{Barnaby:2006km, Mazumdar:2010sa, Enqvist:2004ey, Enqvist:2004bk}.
 In this work we search the parameter space
 of hybrid inflation in details and study the effects of the waterfall
 field dynamics carefully. We show that depending on the model
 parameters inflation can take place at different stages \cite{Randall:1995dj, GarciaBellido:1996qt, Abolhasani:2010kn}.
In conventional hybrid inflation~\cite{Linde:1993cn,Copeland:1994vg}
 inflation ends abruptly by the waterfall transition.
Here we find situations in which inflation can proceed as in chaotic
inflation~\cite{Linde:1983gd} even after the waterfall transition.
This is a new model of double inflation~\cite{Silk:1986vc} 
in the context of hybrid inflation. Interestingly, as we shall see,
 in some situations inflation can actually have three extended stages. 
The key parameter is the ratio $\phi_c/M_P$ where $\phi_c$ is 
the value of the inflaton field at the time of waterfall. 
With  $\phi_c/M_P\gg 1$ one can get a second period of inflation even
 after a sharp waterfall phase transition.

We investigate the curvature perturbations in these models. 
The key constraint is that at the background the waterfall field $\psi$ is
turned off so one has to introduce an effective trajectory
$\sqrt{\langle\delta\psi^2\rangle}$ from the quantum fluctuations of
the waterfall field. This in turn plays crucial roles when one 
calculates the curvature perturbation power spectrum. 
We exclude a large class of model parameters where the curvature
 perturbations cannot be normalized to the COBE/WMAP 
 amplitude~\cite{Bennett:1996ce,Liddle:2006ev}.

The rest of paper is organized as follows. In section \ref{sec:model}
we present our setup. In sections \ref{sec:background}, \ref{sec:curvpert}
 and \ref{sec:corr} we consider a particular model in which inflation 
has three extended stages. After presenting the 
background solutions, we study the waterfall quantum fluctuations
 and the induced  curvature 
perturbations in this model using the $\delta N$-formalism.
 In section \ref{general-limit} we investigate other limits of 
the model parameter space and calculate the curvature perturbations.
 We investigate the model parameters where the induced curvature 
perturbations from the waterfall float between the results obtained 
in sections \ref{sec:background}-\ref{sec:corr} and the results of 
standard hybrid inflation. Some technical calculations of the waterfall 
field dynamics are relegated into 
Appendices \ref{betterapprox}, \ref{curvature-rigorous} and \ref{app:deln-k>kc}.


\section{Hybrid Inflation and its parameters space}
\label{sec:model}

As mentioned above we would like to investigate different limits of  
hybrid inflation potential parameter space and examine the effects of
the waterfall transition on curvature perturbations.
 The potential has the form 
\begin{equation}
\label{pot}
V(\phi,\psi) = \frac{\lambda}{4} 
\left( \psi^2 - \frac{M^2}{\lambda}\right)^2 
+ \frac{1}{2} m^2 \phi^2 + \frac{1}{2}g^2 \phi^2 \psi^2 \, ,
\end{equation}
where $\phi$ is the inflaton field, $\psi$ is the waterfall field, and 
$\lambda$ and $g$ are two dimensionless couplings.
The potential has global minima at $(\phi,\psi)=(0,\pm M/\sqrt{\lambda})$.

In our analysis below, we retain some key properties of standard 
hybrid inflation as proposed in \cite{Linde:1993cn,Copeland:1994vg} 
while relaxing other conditions. More specifically, as in
conventional model of hybrid inflation, we keep the assumption of 
the vacuum domination. That is, during the first stage of inflation
 and before the phase transition starts, inflation is driven basically 
by the vacuum energy $M^4/4 \lambda$. Secondly, we assume that 
the waterfall field is very heavy so it quickly relaxes to its 
instantaneous minimum $\psi=0$ 
before the phase transition. Besides these two key assumptions, 
we relax the other conditions on model parameters. 
For example, in the original version of hybrid inflation $\phi$ is 
sub-Planckian, while here we relax this condition and shall 
assume $\phi_c/M_P$ to be arbitrary where $\phi_c\equiv M/g$ is the
value of the inflaton field at the onset of waterfall.
Furthermore, we keep the ratio $g^2/\lambda$  arbitrary  whereas 
in conventional models of hybrid inflation it is usually assumed 
that $g^2 \sim \lambda$.  
Before we proceed we mention that similar studies have been performed 
in \cite{Clesse:2010iz, Kodama:2011vs}. These works mainly concentrated 
on the classical fields dynamics whereas in our analysis the quantum 
fluctuations of the waterfall field play key roles.

The first period of inflation takes place at $\phi>\phi_c$ and $\psi=0$.
As stressed above, as in a conventional model of hybrid inflation 
we assume this stage is vacuum-energy dominated,
\ba
\frac{M^4}{4 \lambda} \gg \frac{1}{2} m^2 \phi^2\,,
\label{vacdom}
\ea
with the Hubble parameter given by
\begin{eqnarray}
H^2=\frac{V(\phi,0)}{3M_{P}^2}
\simeq\frac{M^4}{12\lambda M_{P}^2}\,,
\label{hubble}
\end{eqnarray}
where $M_{P}=1/\sqrt{8\pi G}$ is the (reduced) Planck mass.
We assume $m^2\ll H^2$ so that $\phi$ is sufficiently slow-rolling
during the first period of inflation before the waterfall. 
 Also, as mentioned before,  we also assume that the waterfall 
field is very heavy so that $\psi=0$ is classically realized before 
the waterfall phase transition. These two assumptions about the masses
 of the inflaton and the waterfall fields can be translated into 
dimensionless parameter $\alpha$ and $\beta$ defined by
\ba
\label{alpha-beta}
\alpha \equiv \frac{m^2}{H^2}\, \quad ,\quad 
\beta \equiv \frac{M^2}{H^2}\,.
\ea
The fact that $\phi$ is slow-rolling implies that $\alpha \ll 1$ 
whereas the assumption that $\psi$ is very heavy requires $\beta \gg1$. 

The condition of vacuum domination, for $\phi$ near $\phi_c$,
can be written as  
\ba
\label{vac-cond1}
\frac{\phi_c^2}{M_P^2} \ll \frac{6}{\alpha} \, .
\ea
Knowing that $\alpha \ll 1$ this constraint indicates that there
 is no sub-Planckian restriction. Indeed one can 
also assume $\phi_c \gg M_P$ and a second stage of inflation may follow 
after a sharp phase transition. Alternatively, one can cast the vacuum
 domination condition in the form
\ba
\label{vac-cond2}
C\equiv \frac{\beta}{\alpha} \frac{g^2}{\lambda} >1 \, .
\ea
This condition indicates a hierarchy of the model parameters required 
to have the vacuum domination. However, the vacuum domination condition 
expressed by Eq. (\ref{vac-cond1}) shed more lights 
because it singles out the ratio $\phi_c/M_P$ as the key parameter 
which effectively distinguishes different limits of hybrid inflation 
as we shall see below.

After $\phi<\phi_c$ the waterfall becomes tachyonic triggering an 
instability in the system. 
We assume that this phase transition is mild enough such that
it will take a few tens of $e$-folds for the waterfall field to 
settle down to its global minimum. Again this is possible if $\phi$ 
stays near the critical point for a sufficiently long time. 
As we shall see below this condition requires $m^2M^2\ll H^4$.
After the waterfall transition the quantum fluctuations
of $\psi$ becomes tachyonic and will grow exponentially.
The accumulative effects of 
these growing modes, i.e. $\langle\delta\psi^2\rangle$,
through the interaction term changes the dynamics of the system.
Soon after the gradient term in the waterfall field dynamics becomes
negligible compared to the effective mass term the quantum
 fluctuations $\delta \psi$ obey the same equation as the background
homogeneous $\psi$ equation.
Hence they effectively behave like a homogeneous classical background,
$\psi_{\rm classical}=\sqrt{\langle\delta\psi^2\rangle}$, however see 
also \cite{Lyth:2010zq}.

The tachyonic growth of $\delta \psi$
continues until the self interaction term $\lambda \psi^4$
becomes important. We call this stage the waterfall stage. It
ends when $\psi$ settles to its global minimum 
$\psi_{min} = \frac{\pm M}{\sqrt \lambda}$ and the symmetry breaking 
is completed. The other important effect of the 
tachyonic growth of $\delta \psi$ is its back-reaction 
on the inflaton field via the interaction $g^2 \phi^2 \psi^2$.
This can increase the effective mass of the inflation field 
significantly ending its slow-roll sharply.
The key question is which of the two back-reactions, 
$\lambda \psi^4$ or $g^2 \phi^2 \psi^2$, becomes important sooner. 
In standard hybrid inflation with $\beta$ extremely large, 
say $\beta \sim 10^3$, $\phi_c\ll M_P$ and $g^2 \sim \lambda$ these
 two back-reactions become important more or less at the same time.
In this view the time of end of inflation is nearly the same as the 
time when the symmetry breaking is complete. 

In this work we distinguish the completion of symmetry breaking
and the termination of inflation as two different phenomena.
We will show that, depending on the model parameters, one can consider 
situation where the waterfall stage and symmetry breaking completion
 happens at an early stage of inflation, followed by a long period of 
the second stage of inflation. As we shall see the key parameter 
in the analysis is the ratio $\phi_c/M_P$.

The effective mass of the inflaton can be read as
\ba
\label{m-eff}
m^2_{eff}= m^2 + g^2 \langle\delta\psi^2\rangle
\ea
Knowing that $m^2 \ll H^2$ (or equivalently $\alpha \ll 1$),
inflation can only end when 
$g^2\langle\delta\psi^2\rangle\simeq H^2\simeq M^4/12\lambda M_P^2$.
Denoting this value of $\psi$ by $\psi_{trig}$ which triggers 
the fast roll of $\phi$ towards its minimum, we find
\ba
\psi^2_{trig} \simeq \frac{M^4}{12 \lambda g^2 M_P^2}
 = \frac{\phi_c^2}{12 M_P^2} \psi_{min}^2 \, ,
\ea
where  $\psi_{min} = \frac{\pm M}{\sqrt \lambda}$.
 For $\phi_c /M_P < \sqrt {12}$, one has 
$\psi_{trig} < \psi_{min}$ so the back-reaction of $\psi$ on $\phi$ 
becomes important sooner than the self-interaction $\lambda \psi^4$
 so inflaton field rolls quickly to its minimum before 
the symmetry breaking completion. During this transition, 
inflation may proceed briefly via the remaining vacuum energy 
but inflation will end quickly once the waterfall field nearly 
settles down to its minimum. 
This is the case in conventional hybrid inflation. On the other hand, 
for  $\phi_c /M_P > \sqrt {12}$ one has $\psi_{trig} > \psi_{min}$,
 meaning that the self interaction $\lambda \psi^4$ becomes 
more important than the back-reaction $g^2 \phi^2 \psi^2$.
 As a result, the waterfall field settles down to its minimum while 
inflation still proceeds afterwards as in chaotic inflation. 
Of course the dynamics of symmetry breaking is a violent phenomena
 which can cause a sudden violation of the slow-roll conditions. 
However, due to the attractor nature of the chaotic inflation potential,
the slow-roll conditions are restored quickly for the second period
of inflation.

In our analysis it is convenient to use the number of $e$-folds
 $d N= H dt$ as a clock. For convenience we set $N=0$ at the time
when the comoving scale corresponding to the present Hubble radius
crossed the horizon during inflation.
We denote the number of $e$-folds till the time 
when the waterfall field becomes unstable at $\phi=\phi_c$ by $N=N_c$.
Taking this epoch as the reference point 
we also define $n\equiv N- N_c$, so for the period before phase 
transition (after phase transition) one has $n<0 (n>0)$. 

\subsection{$\phi_c^2 > 12 M_P^2$}
For $\phi_c^2 > 12 M_P^2$, one has $\psi_{trig} > \psi_{min}$ and
the back-reaction of $\psi$ on itself becomes important first. 
The time when it takes for the waterfall field to settles 
to its global minimum depends on the sharpness of the phase transition.
Denote by $n=n_p$ the number of $e$-folds it take for the waterfall
field to settle down to its minimum. For a very sharp phase transition,
$n_p \lesssim1 $ whereas for a mild phase transition $n_p \gg 1$.

For $n> n_p$, the effective potential for inflaton field is 
$V_{eff}= \frac{1}{2}m_{eff}^2 \phi^2$ with $m_{eff}$ given by
 Eq.~(\ref{m-eff}). For $\phi_c^2 > 12 M_P^2$ this leads to a period 
of chaotic inflation where $\phi$ evolves as 
\ba
\phi^2{(n)} = \phi_p^2 - 4 M_P^2 (n-n_p) \, ,
\ea
where $\phi_p$ indicates the value of $\phi$ at the time of $n_p$.
As in chaotic inflation, inflation ends at $\phi=\phi_f$ where 
the slow-roll conditions are violated. The number of $e$-folds 
during the chaotic era, $\Delta n_{ch}$, is 
\ba
\Delta n_{ch}  = \frac{\phi_p^2 - \phi_f^2}{4 M_P^2}
\ea
The total number of $e$-folds, therefore, is
\ba
\label{N-tot}
N_{tot} = N_c + n_p + \Delta n_{ch} \, .
\ea 
Here we have called the number of $e$-folds from $N=0$ until the
end of inflation the total number of $e$-folds for convenience,
but it should be noted that this number is nothing to do with
the actual total number of $e$-folds of inflation which may well
be almost infinite.

To solve the flatness and the horizon problem we require $N_{tot}\gtrsim 60$.
This picture only depends on the assumption that $\phi_c^2 > 12 M_P^2$
and is independent of the values of $\beta$ and the ratio of the
 couplings $g^2/\lambda$.
As we shall see below, the sharpness of the phase transition
 (the magnitude of $n_p$) depends on the parameter
 $\kappa_\psi  \simeq  \alpha \beta/3 $. If $\kappa_\psi  \ll1$ then 
phase transition is mild and $n_p>1$. In this limit, as can be seen
 from Eq.~(\ref{N-tot}), inflation has three stages. 
For this to happen, knowing that $\alpha \sim 1/N_{tot} $, 
one requires that  $1 \ll \beta \ll N_{total}$ so  $\beta \sim 10$ or so.
On the other hand, if $\kappa_\psi  >1$ then the phase transition is 
very sharp and $n_p\lesssim 1$.  However, due to smallness of $\alpha$, 
to have $\kappa_\psi $ one requires that $\beta \gg  N_{tot}$,
 i.e. $\beta \sim 10^2-10^3$.

\subsection{$\phi_c^2 < 12 M_P^2$}

In this limit, $\psi_{trig}^2 < \psi_{min}^2 $ so the back-reaction
 $g^2 \psi^2 \phi^2$ becomes 
important sooner than the self-interaction $\lambda \psi^4$, 
triggering the inflaton field to roll down quickly to its 
 minimum $\phi=0$. Thus inflation cannot continue for
 an extended period in the form of chaotic inflation,
and one expects that the completion of symmetry breaking
and the end of inflation occurs more or less at the same time.
Specifically
\ba
N_{tot}= N_c + n_{trig} + \Delta n_{\phi-trans}  +   \Delta n_{\psi-trans} \, .
\ea
Here $n_{trig}$ denotes the time when the back-reaction of $\psi$ on the
inflaton mass become significant, triggering its fast rolling to $\phi=0$, 
whereas $\Delta n_{\phi-trans} $ denotes the number of $e$-folds it takes 
for the inflaton field to move toward its minimum until the time when
the coupling term $g^2 \phi^2 \psi^2$ becomes negligible compared to
the waterfall bare mass term $-M^2\psi^2$, i.e. until
when $\phi^2\ll\phi_c^2$. Since $\phi$ evolves like $e^{-3n/2}$ 
this period is rather short, $\Delta n_{\phi-trans} \sim 1$. 
Once the value of $\phi$ becomes negligible, the $\psi$ field becomes 
even more tachyonic and rolls quickly dow to its minimum like 
$\psi \simeq \psi_{trig} e^{\sqrt \beta (n-n_{trig})}$.

For large $\beta$ this period of the $\psi$ transition
 (symmetry breaking completion) is very quick,
$\Delta n_{\psi-trans} \ll 1$. One can show that 
$n_{trig} \sim \Delta n_{\psi-trans} \sim \frac{1}{\sqrt \beta} <1$,
 so inflation ends abruptly after the waterfall phase transition
 as in standard hybrid inflation.

Having specified two major scenarios of hybrid inflation based on
 the ratio $\phi_c/M_P$, we now proceed to the study of curvature 
perturbations in these scenarios. The case with $\phi_c^2 < 12 M_P^2$
 is somewhat similar to standard hybrid inflation as studied 
in \cite{Lyth:2010ch,Abolhasani:2010kr,Fonseca:2010nk,Gong:2010zf,Lyth:2010zq}.
We shall come back to this limit in section \ref{general-limit}. 
The case with $\phi_c^2 > 12 M_P^2$ is the one that needs 
careful considerations. In the next three sections, as a sample
 example, we consider in details a model with 
$\phi_c^2 > 12 M_P^2$ and  $\alpha \beta \ll 1$ so $n_p \gg1$ 
and inflation has three extended stages. This case perhaps is 
the most elaborate model in our case study which would be helpful
 in the classification of hybrid inflation
models in section \ref{general-limit}.
Once we have performed the curvature perturbation analysis in this case, 
we are able to make a connection to the case $\phi_c^2 < 12 M_P^2$ and
 provide a global view of the issues of curvature perturbations 
 and the waterfall dynamics in different limits of 
the parameter space of hybrid inflation.

\section{Background Dynamics}
\label{sec:background}

The analysis performed in this section and sections \ref{sec:curvpert} 
and \ref{sec:corr} are for the case $\phi_c^2 > 12 M_P^2$ and
 $\alpha \beta \ll 1$. In this section we study the background 
dynamics in details which is also applicable to other cases.  

We assume the standard metric,
\ba
ds^2 = - dt^2 + a(t)^2 d{\bmx}^2 \, ,
\ea
where $a(t)$ is the scale factor. As mentioned before, we
assume the vacuum energy dominance for most of the inflationary
era. Hence we have
\begin{eqnarray}
a(t)\propto e^{Ht}\,,
\end{eqnarray}
where $H$ is given by (\ref{hubble}).

In this approximation, the dynamics of the system is solely
described by the field equations for $\phi$ and $\psi$.
 The field equations for the classical, homogeneous background 
fields are written as 
\ba
\label{phi-c}
\phi'' + 3  \phi'
 + \left(\alpha +g^2  \frac{\psi^2 }{H^2} \right) \phi =0\,,
\\
\label{psi-c}
\psi'' + 3  \psi' 
+ \left(-\beta +g^2 \frac{\phi^2}{H^2}
+\lambda \frac{\psi^2} {H^2}\right) \psi =0 \, ,
\ea
where the prime denotes the differentiation with respect to $n$,
${}'=d/dn$, where $n=N-N_c$, and the dimensionless parameters 
$\alpha$ and $\beta$ are introduced in Eq. (\ref{alpha-beta}).
We assume $\alpha\ll1$ and $\beta \gg 1$. 
Here and in the following, we denote the classical fields by 
$\phi$ and $\psi$, while the quantum 
fluctuations by $\delta\phi$ and $\delta\psi$.

At early times $(-n)\gg1$, the classical waterfall field is trapped
at $\psi=0$ and the quantum fluctuations are completely negligible.
Hence if we simply solve the classical field equation (\ref{psi-c}) with
the initial condition $\psi=\psi'=0$,
$\psi$ would stay at the origin forever even at $n>0$.
However, because of the exponential growth of the 
quantum fluctuations of the waterfall field after the transition,
the expectation value $\langle\delta\psi^2\rangle$ becomes non-negligible
and the rms value soon starts to behave as a classical field
$\psi=\sqrt{\langle\delta\psi^2\rangle}$. In fact we will show in the
next subsection that $\langle\delta\psi^2\rangle$ is dominated by
the superhorizon modes, hence the rms of it is indeed observed as
a classical background for an observer within each Hubble horizon region.
In particular, in Eqs.~(\ref{phi-c}) and (\ref{psi-c}),
$\psi^2$ in the interaction terms $g^2\psi^2/H^2$ and $\lambda\psi^2/H^2$
are to be considered as $\langle\delta\psi^2\rangle$.

Here we provide the solutions for the $\phi$ and $\psi$ evolution,
starting with the dynamics of $\phi$.
Let us first assume that the interaction term $g^2 \psi^2\phi^2$
is negligible. Namely, we consider the stage when 
$\alpha\gg g^2\psi^2/H^2$ in Eq.~(\ref{phi-c}).
With this assumption one can easily solve it to obtain
\ba
\label{phif}
\phi(n) =\phi_c \exp \left(-r\,n \right)\quad(n<n_p)\,,
\ea
with
\ba
r \equiv \left( \frac{3}{2}- \sqrt{\frac{9}{4}-\alpha}\right) 
\simeq \frac{\alpha}{3} \, .
\ea
This stage ends when the value of $g^2\psi^2/H^2$ becomes comparable
to $\alpha$. To simplify the dynamics, we assume that this
happens at about the end of the waterfall stage, $n=n_p$,
when $\psi^2=\langle\delta\psi^2\rangle$
levels off to the local minimum.

Now we look into the dynamics of $\psi$. 
Before the transition $\psi=0$ but it effectively becomes
non-vanishing after the transition.
During the waterfall stage, we can neglect
the self-interaction $\lambda\psi^4$,
and the equation for $\psi$ simplifies to
\ba
\label{psi-substituted}
\psi''+3\psi' +\beta \left( e^{-2r\,n} -1\right)\psi =0 \, .
\ea
The general solution is given by a linear
combination of the Bessel functions $J_\nu(z)$ and $Y_\nu(z)$,
where $\nu=\frac{\sqrt{\beta+9/4}}{r}$ and $z=\frac{\sqrt{\beta}e^{-rn}}{r}$.
However, one can get a better view of the $\psi$ solution using the
WKB approximation as we demonstrate below. 

Defining  $\psi = \Psi e^{-3n/2}$, Eq.~(\ref{psi-substituted}) 
is rewritten as
\ba
\Psi'' +\left(-\frac{9}{4} -2 \beta r\,n+2 \beta r^2\,n^2 \right) \Psi =0 \, ,
\ea
in which we expanded the exponential in Eq.~(\ref{psi-substituted})
to second order in $rn \ll 1$. Using the WKB approximation one has
\ba
\label{psi-int}
\Psi \sim \exp 
\left[\int dn \sqrt{\frac{9}{4}+2 \beta r\,n-2 \beta r^2\,n^2}\right] \,.
\ea 

As can be seen from Eq. (\ref{psi-substituted}) (see also
 Eq. (\ref{deltapsi-de}) below) the sharpness of the phase transition
is controlled by the product $\beta r$. The larger is $\beta r$ the greater
is the effective tachyonic mass of the waterfall during the phase
transition, causing a sharp phase transition. This means that,
by keeping the product $\beta r$ small one basically reduces the
effective tachyonic mass of the waterfall field during the phase 
transition so inflation can continue for an extended period during 
the course of the phase transition.
In the analysis below, we are interested in a situation where 
$\beta r \ll1$ so the phase transition is not sharp and $n_p \gg 1$.
 This is a new limit where the curvature perturbation shows some 
interesting behavior. However, for this to happen one requires 
$\beta \ll 1/r$ so $\beta$ can not be arbitrarily large. 
For reasonable values of $r$ this requires $\beta \sim 10$. 
On the other hand, if one is interested in the limit where 
$\beta r >1$, corresponding to $n_p \sim 1$,
then the analysis of the curvature perturbation from the waterfall is 
basically the same as in standard hybrid inflation. We will 
comment on this case later. In our analysis it is useful to define
\ba
\label{epsioln-psi}
\kappa_\psi  \equiv \frac{\beta\, r}{3}\simeq\frac{\alpha\beta}{9} \ll1\,.
\ea

\begin{figure}
\centerline{\includegraphics[scale=.4]{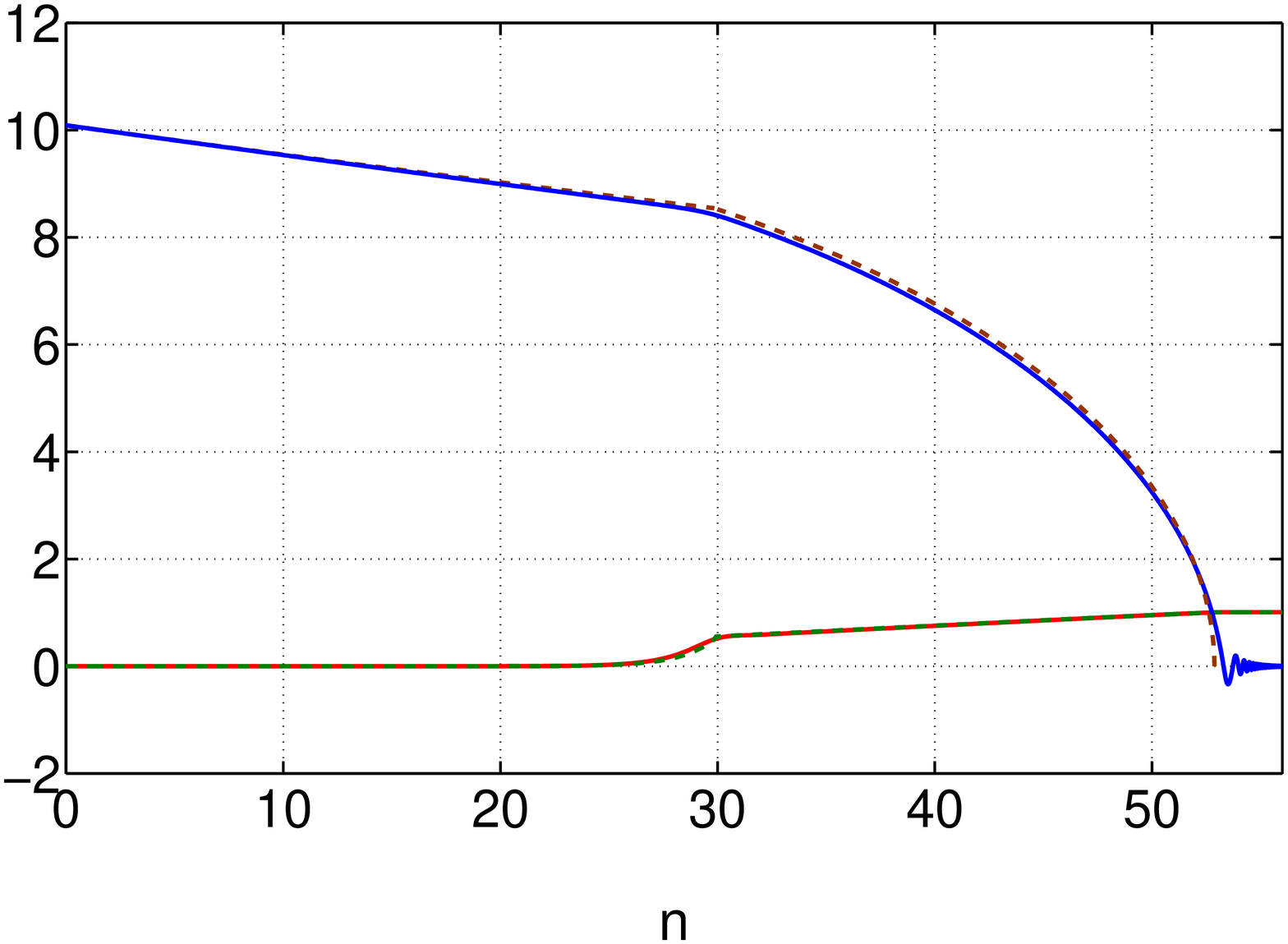}}
\caption{
Here we present the background fields dynamics in units of $\mpl$ from the
phase transition $n=0$ to the end of inflation. The solid blue line and 
the solid red line, respectively,  show the full numerical solutions
 of the $\phi$ and $\psi$ evolutions with the 
initial value  $\psi(n=0) = \langle \delta \psi^2 (0)\rangle$. 
The dashed green line and the dashed brown line, respectively, 
show our approximate analytical solutions for the corresponding fields.
The agreement between them is very good. In this picture, 
the time of phase transition completion is $n_p\simeq 30$.
The numeric analysis are for the parameters,  
$M=5 \times 10^{-5} \mpl$, $m=2 \times 10^{-6} \mpl$, 
$g^2 =2 \times 10^{-11}$ and $ \lambda = 2 \times 10^{-9}$
corresponding to $\beta \simeq 8.5$ and $ \kappa_\psi  \simeq 0.02$. } 
\vspace{0.7cm}
\label{back-fig} 
\end{figure}

Calculating the integral in Eq.~(\ref{psi-int}) to leading orders in
$\kappa_\psi $  and setting the initial condition at the moment of
the phase transition $n=0$ yields
\ba
\label{psiapp}
\psi(n)= \psi(0) 
\exp\left[\kappa_\psi  n^2 -\frac{4}{9}\kappa_\psi ^2 n^3\right] \, ,
\ea
where  $\psi(0)=\sqrt{\langle\delta\psi^2(0)\rangle}$.
The good agreement between this solution and the full numerical 
solution can be seen in Fig.~\ref{back-fig}.  Note that the term 
in Eq.~(\ref{psiapp}) containing $\kappa_\psi ^2 n^3$ is much 
smaller than the term $\kappa_\psi  n^2$ 
so we shall safely ignore it in most of our analysis below.

The above solution is valid until the self interaction term 
$\lambda\psi^4$ becomes important.
 Eventually the self-interaction becomes important and
$\psi$ settles down to its local minimum because of the large 
induced mass, $\partial_\psi^2V(\psi_{min})=2(M^2-g^2\phi^2)\simeq2M^2$. 
Here note that the value of $\phi$ will be a fraction of $\phi_c$
by the time $\psi^2$ reaches $\psi_{min}^2(\phi)$ (see the next paragraph)
and hence we may assume $M^2\gg g^2\phi^2$. Denoting the time at which
$\psi$ reaches the local minimum by $n=n_p$ we have
\begin{eqnarray}
\label{psi-min-phi}
\psi^2(0)
\exp\left[2\kappa_\psi  n_p^2-\frac{8}{9} \kappa_\psi ^3 n_p^3\right]
=\psi_{min}^2(\phi)=\frac{M^2}{\lambda}-\frac{g^2\phi^2}{\lambda}
\simeq\frac{M^2}{\lambda}\,,
\end{eqnarray}
or to leading order 
\begin{eqnarray}
n_p^2=-\frac{1}{2\kappa_\psi }
\ln\left[\frac{\lambda\,\psi^2(0)}{M^2}\right]\,.
\end{eqnarray}
This equation tells us how the total number of $e$-folds
of the waterfall stage depends on $\psi^2(0)$.
As discussed in detail in the next section,
it will play the central role in the
evaluation of the curvature perturbation.

Now, since we have assumed that $g^2\psi^2/H^2$ becomes
comparable to $\alpha$ at the end of the waterfall stage,
we should examine its effect on the evolution of $\phi$.
Inserting the solution (\ref{psiapp}) for $\psi$ into
Eq.~(\ref{phi-c}), one finds instead of Eq.~(\ref{phif}),
\begin{eqnarray}
\phi(n)=\phi_c\exp\left[-rn-\frac{g^2\psi^2(0)}{3H^2}\int_0^n
e^{2\kappa_\psi  n'{}^2}dn'
\right]\,.
\label{phiwfstage}
\end{eqnarray}
At the end of the waterfall stage, $n=n_p$,
we have $\psi^2(n_p)=\psi_{min}^2(\phi)\simeq M^2/\lambda$.
Hence the above gives
\begin{eqnarray}
\phi(n_p)=\phi_c\exp\left[-rn_p-\frac{g^2\beta}{3\lambda}\right]
=\phi_c\exp\left[-\frac{\alpha}{3}n_p-\frac{\alpha}{3}C\right]
\,,
\end{eqnarray}
where we have used Eq.~(\ref{vac-cond2}) in the second equality.


After $\psi^2$ settles down to $\psi_{min}^2(\phi)$, i.e. at the
stage $n>n_p$, we can plug this minimum value into the potential 
to obtain the effective potential for $\phi$,
\ba
\label{eff-pot-phi}
V_{eff}(\phi) &=&V\bigl(\phi,\psi_{min}(\phi)\bigr)
=  \frac{1}{2}m^2 \phi^2 
+ \frac{g^2}{2 \lambda} M^2 \phi^2 -\frac{g^4}{4 \lambda} \phi^4 
\nonumber\\
&\simeq& 
\frac{1}{2}\left[m^2 + \frac{g^2}{\lambda} M^2\right] \phi^2
\,,
\ea
where the second line follows from our assumption that
we have $M^2> g^2\phi^2$ by the end of the waterfall stage.
Hence the $g^2\phi^2/2$ term may be ignored at the stage
$n>n_p$. The effective mass of $\phi$ at the stage $n>n_p$ is given by
\ba
m^{2}_{\phi,eff} =m^2+ \frac{g^2}{\lambda} M^2 = m^2 (1+C) \, .
\ea
We note that the potential now behaves exactly the same as in 
standard chaotic inflation with  $m^2 \rightarrow m^2 (1+C)$. 
In other words, the universe is no longer vacuum-dominated at
 $n>n_p$ and inflation proceeds as in chaotic inflation.

By using the above form of the effective potential, 
and assuming $\phi(n_p)\gg M_{P}$,
the equation for $\phi$ at $n>n_p$ reduces to 
\ba
-8 M_{pl}^2 (n- n_p) = ( \phi^2 - \phi_p^2) - \left( 1+ C^{-1} \right) \phi_c^2
\ln  \left[ \frac{1- \frac{\phi_p^2}{\phi_c^2}
\left( 1+ C^{-1} \right)^{-1}}{1- \frac{\phi^2}{\phi_c^2} 
\left( 1+ C^{-1} \right)^{-1}} \right]\, ,
\ea
which for $\phi, \phi_p< \phi_c$  and $C > 1$  can be approximated to 
\ba 
\phi^2(n) =\phi_p^2-4M_{pl}^2(n-n_p)\quad (n>n_p)\,,
\label{phiend}
\ea
where $\phi_p\equiv\phi(n_p)=\phi_ce^{-rn_p}$. The slow-roll inflation
ends when $\phi$ becomes of order $M_{P}$ as usual (approximately at
$\phi=\phi_f\simeq\sqrt{2/3}\,M_{P}$). Here it should be stressed
that the end of inflation is solely determined by the value
of $\phi$ alone. This is an important point to be kept in mind
when calculating the curvature perturbation using the 
$\delta N$-formalism~\cite{Sasaki:1995aw}
 (see discussions in Sec.~\ref{sec:deltaN}).
The overall evolution of the classical fields is
depicted in Fig.~\ref{back-fig}.

It is worth noting again that the above evolutionary
picture holds at any horizon patch of the universe.
Until the time of transition, $\psi=\sqrt{\langle\delta\psi^2\rangle}$
cannot be regarded as classical. But as soon as the tachyonic instability
sets in, it starts to behave like a classical field.
Another important point is that the initial amplitude of these
 fluctuations at $n=0$ varies from one patch to another because of
the very nature of the quantum fluctuations. As will be discussed
in Sec.~\ref{sec:curvpert} below, this implies the spatial variation
of the number of $e$-folds from the time of transition 
to the end of the waterfall stage $n=n_p$, which is fixed
by the condition $\langle\delta\psi^2(n_p)\rangle=\psi_{min}^2(\phi)$,
giving rise to the curvature perturbation at the end of inflation.
It may be also noted that the quantum fluctuations of
the waterfall field play the role of entropy perturbations
if we regard $\phi$ as the inflaton field.
However, as we have described in the above, since they
give rise to a non-vanishing classical background field $\psi$
whose evolution is crucial to the dynamics of inflation,
the system is better regarded as a two-field inflation model
after the waterfall transition. 

\section{Curvature perturbation from the Water-fall Field}
\label{sec:curvpert}
In this section we study the quantum fluctuations of the waterfall
field $\delta\psi$ and calculate the curvature perturbation from
them by using the $\delta N$-formalism. 


\subsection{Quantum fluctuations}
\label{sec:quantum}
The equation governing fluctuations of 
the waterfall field in real space is
\ba
\label{psi-eq-conf}
\delta \psi'' + 3 \delta \psi'
- \frac{1}{a^2H^2} \nabla^2 \delta \psi 
+\left( -\beta +g^2 \frac{\phi^2}{H^2} 
 +3\lambda \frac{\psi ^2}{H^2} \right)
 \delta \psi =0 \, ,
\ea
where $\psi$ is the classical background.
As we discussed in the previous section, 
we eventually identify it as $\psi^2=\langle\delta\psi^2\rangle$.
At early times before the transition or even after the transition, 
the self interaction term $\lambda \langle \delta \psi^2\rangle/{H^2}$
in the mass term of Eq.~(\ref{mspace}) is negligible until the
end of the waterfall stage. In this approximation Eq.~(\ref{psi-eq-conf})
in momentum space $\bmk$ is given by
\ba
\delta \psi_{\bm{k}}'' + 3 \delta \psi_{\bm{k}}'
 + \left( \frac{k^2}{a^2 H^2}
 -\beta +g^2 \frac{\phi^2}{H^2} \right)
 \delta\psi_{\bm{k}}=0 \, ,
\label{mspace}
\ea
where 
\begin{eqnarray}
\delta\psi_{\bm k}
=\int \frac{d^3x}{(2\pi)^{3/2}}
\delta\psi({\bm x})e^{-i{\bm k}\cdot{\bm x}}
=\left(a_{\bmk}\psi_k(n)
+a_{-\bmk}^\dag\overline{\psi_k(n)}\right)\,.
\end{eqnarray}
Here $a_{\bmk}$ and $a_{\bmk}^\dag$ are the annihilation
and creation operators, repsectivly, with respect to
a suitably chosen vacuum and $\psi_k(n)$
is the positive frequency function. 

Substituting the solution~(\ref{phif}) for $\phi$, one obtains
\ba
\label{deltapsi-de}
\delta \psi_{k}'' + 3 \delta \psi_{k}' 
+ \left( \frac{k^2}{k_c^2}e^{-2n}
 + \beta \left( e^{-2 r n}-1 \right) \right) \delta \psi_{k} =0 \,,
\ea
where $k_c$ is the comoving wavenumber that crosses the horizon
at the critical point $n=0$, $k_c=Ha(n=0)$.

We assume a sufficiently small $r$ such that $r|n|\ll 1$ 
for all scales $k$ of cosmological interest. Hence we 
have $\beta|e^{-2rn}-1|\simeq 2\beta r |n|$, which is always smaller than unity 
because of our assumption that $\kappa_\psi  \ll1$.
Hence when the wavenumber $k$ is inside the horizon, $k>k_ce^{n}$,
one can neglect the term proportional to $\beta$.
The general solution in this limit is given in terms of 
the Hankel functions $ H^{(1)}_{3/2} (x)$ and $ H^{(2)}_{3/2} (x)$
by
\begin{eqnarray}
\delta \psi_{k}=e^{-3n/2}
\left(c_1 H^{(1)}_{3/2} (x)+c_2 H^{(2)}_{3/2}(x)\right)\,;
\quad x\equiv \frac{k}{k_c}e^{-n}\,.
\end{eqnarray}
As usual, the natural choice of the vacuum is that it approaches
the Minkowski mode function in the short distance limit,
\ba 
\delta \psi_k \to \frac{e^{-ik \tau} }{a \sqrt{2k}}  \qquad \mathrm{as} 
\quad k\tau \to -\infty \, ,
\ea
where $\tau$ is the conformal time, $d\tau=dt/a$.
With this initial condition the positive frequency
function for the modes inside the horizon is obtained to be 
\begin{eqnarray}
\label{deltapsi-}
\delta \psi_k(n)
=\sqrt{\frac{\pi H^2}{4 k_c^3}}\,e^{-3n/2}
H^{(1)}_{3/2} \left( \frac{k}{k_c}~e^{-n} \right)  \, .
\end{eqnarray}
At the time of horizon crossing $n=n_k$ when $e^{n_k}=k/k_c$,
the above expression gives
\ba
\left \vert \delta \psi _{k}(n_k)\right \vert \simeq 
\frac{H}{\sqrt{2k^3}} \, .
\label{psihc}
\ea

\begin{figure}
\centerline{\includegraphics[scale=.4]{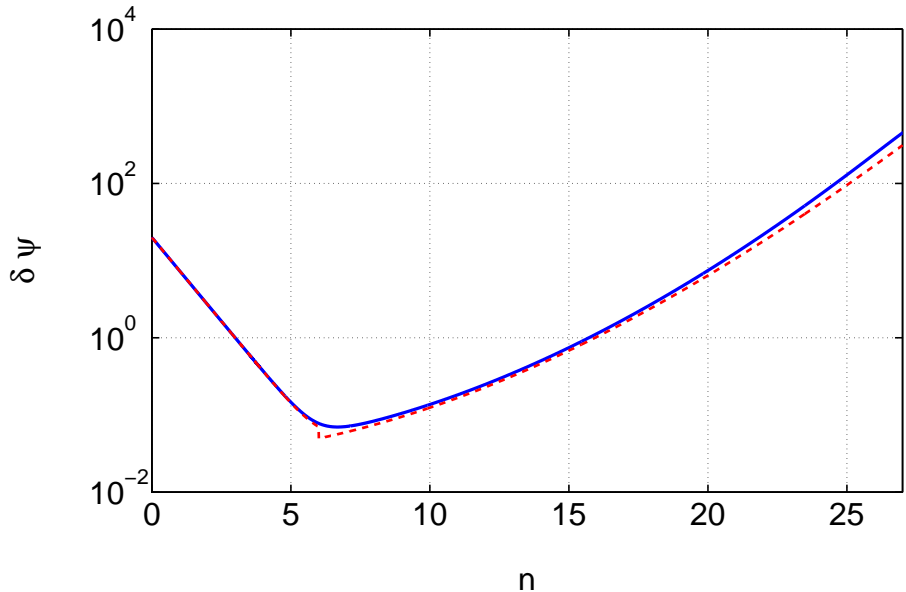}}
\caption{
The amplitude of the mode function $|\psi_{k}(n)|$ as a function of 
the number of e-folds $n$ is shown for a mode which leaves the horizon
 at $n=6$. The solid blue line is obtained from the full numerical analysis
 whereas the dashed red line is from our analytical 
expression~(\ref{dpsibetter}). All numerical parameters are the same as
in Fig.~\ref{back-fig}.
} 
\vspace{0.7cm}
\label{qpsi} 
\end{figure}

After horizon crossing one can neglect the spatial
gradient term $(k/k_c)^2e^{-2n}$
in Eq.~(\ref{deltapsi-de}) and the equation for $\psi_k$
 becomes identical to that for the classical background, 
Eq.~(\ref{psi-substituted}). Hence the solution 
takes the form~(\ref{psiapp}). Matching it with (\ref{psihc})
at $n=n_k$, we obtain an approximate expression for $\psi_k$ at
$n>n_k$ as 
\ba
\delta \psi_k(n)=\frac{H}{\sqrt{2 k^3}}
 \exp\left[\kappa_\psi(n^2 - n_{k}^2)
-\frac{4}{9}\kappa_\psi ^2(n^3-n_k^3)\right]\,.
\label{dpsibetter}
\ea
One can see a good agreement of this analytical expression
with the numerical result in Fig.~\ref{qpsi}. 

\subsection{Waterfall stage}
\label{waterfallstage}

Given the solution to the mode functions, an important quantity
to compute is the squared fluctuation $\delta\psi^2$, because
it determines the duration of the waterfall stage. 
It is given by
\ba
&&\delta\psi^2(n,{\bm x})
=\delta\psi^2(0,{\bm x})
\exp\left[2 \kappa_\psi n^2-\frac{8}{9}\kappa_\psi ^2n^3 \right]\,;
\cr\cr
&&\quad
\delta\psi^2(0,{\bm x})
=\iint_{k,k'<aH} 
\frac{d^3k\,d^3k' }{(2\pi)^{3}}
\delta \psi_{\bm k}(0)\delta\psi_{{\bm k}'}^\dag(0) 
e^{i({\bm k}-{\bm k}')\cdot\bm{x}}\,,
\label{del-psi0-def}
\ea
where
\begin{eqnarray}
\delta \psi_{\bm k}(0)
=\psi_k(0)A_{\bm k}
=\frac{H}{\sqrt{2k^3}}e^{-\kappa_\psi  n_k^2+\frac{4}{9}\kappa_\psi  n_k^3}
A_{\bm k}\,;
\quad A_{\bm k}\equiv\left(a_{\bm k}+a_{-\bm k}^\dag\right)\,,
\label{spdpsi}
\end{eqnarray}
and we have absorbed an irrelevant phase factor in the
definitions of $a_{\bm k}$ and $a_{-\bm k}^\dag$.
Note that the integral is cut off at the horizon scale because
the modes on subhorizon scales are just the standard vacuum 
fluctuations which are to be regularized to zero. 

The above expression for $\delta\psi_{\bmk}$ shows that
it is a classical random fluctuation, because it is proportional
to $A_{\bmk}$ which commutes with $A_{\bmp}^\dag$ ($=A_{-{\bmp}}$)
for any pair of ${\bmk}$ and ${\bmp}$.
 Hence for each horizon-size patch 
one would observe $\delta\psi^2$ as a homogeneous classical background
which varies smoothly over scales larger than the horizon scale.
Thus on a given, sufficiently large scale, say the comoving scale of 
the present Hubble horizon size, one can calculate the mean value
$\langle\delta\psi^2(n) \rangle$ and the fluctuation
\ba
\label{Delta-def}
\Delta \psi ^2 (n,\bm{x})
\equiv \delta \psi ^2 (n,\bm{x})- \langle \delta \psi^2(n) \rangle\,,
\ea
where $\langle\delta\psi^2(n) \rangle$
determines the homogeneous background while $\Delta \psi ^2 (n,\bm{x})$
gives rise to the curvature perturbations on superhorizon scales.

Let us first consider the background dynamics.
Taking the expectation value of $\delta\psi^2$, we find
\begin{eqnarray}
\langle\delta\psi^2(n)\rangle
=\langle \delta \psi^2(0) \rangle
\exp[2\kappa_\psi  n^2-\frac{8}{9}\kappa_\psi ^2n^3]\,,
\end{eqnarray}
where
\begin{eqnarray}
\langle \delta \psi^2(0) \rangle=
\int_{k<aH}\frac{d^3k}{(2\pi)^3}|\psi_k(0)|^2
=\left(\frac{H}{2\pi}\right)^2
\int_{k<aH}\frac{dk}{k}\exp[-2\kappa_\psi  n_k^2+\frac{8}{9}\kappa_\psi^2n_k^3]\,.
\end{eqnarray}
By changing the variable of integration to $n_{k}=\ln(k/k_c)$,
we may approximate the upper and lower limits of the integral to
plus and minus infinity, respectively, for $\kappa_\psi  n^2\gg1$,
and neglect the term $\kappa_\psi ^2n_k^3$ in the exponent,
to obtain
\ba
\label{del2-ave}
\langle \delta \psi^2(0) \rangle \simeq
 \left(\frac{H}{2\pi}\right)^2
 \int _{-\infty}^{\infty}dn_{k}\exp\left( -2\kappa_\psi n_{k}^2\right)
=\frac{H^2}{4 \pi^2} \sqrt{\frac{\pi}{2 \kappa_\psi}}\,.
\ea
The waterfall stage ends when the self-interaction term becomes
important and $\langle\delta\psi^2\rangle$ reaches the local minimum
$\psi_{min}^2(\phi)$ given by Eq.~(\ref{psi-min-phi}). Thus we have
\begin{eqnarray}
\langle\delta\psi^2(n_p)\rangle
=\langle \delta \psi^2(0) \rangle
\exp[2\kappa_\psi  n_p^2-\frac{8}{9}\kappa_\psi ^2n_p^3]
=\psi_{min}^2(\phi)\,.
\end{eqnarray}
Neglecting the $\kappa_\psi ^2n_p^3$ term, we may approximately
solve this for $n_p$ to find
\ba
{n}_p^2 \simeq -\frac{1}{2 \kappa_\psi}
 \ln \left(\frac{\langle \delta \psi^2(0) \rangle}
{\psi_{min}^2(\phi)}\right) 
\simeq -\frac{1}{2 \kappa_\psi}  \ln \left( \frac{\beta}{\lambda} \sqrt{32 \pi^3 \kappa_\psi }
\right) \, .
\label{np-av}
\ea

Before closing this subsection,
let us make an important comment concerning the meaning of
Eq.~(\ref{spdpsi}).
If we would literally consider it as $\delta\psi^2(n,{\bm x})$ at $n=0$,
then it could only take account of the comoving wavenumbers $k$ smaller
than $k_c$. However, as can be seen from Eq.~(\ref{del-psi0-def}),
it takes account of all the comoving wavenumbers which has crossed
the horizon at $n=n_k$. This implies that $\delta\psi^2(0,{\bm x})$
introduced in Eq.~(\ref{spdpsi}) is defined just for convenience
to express $\delta\psi^2(n,{\bm x})$ at the end of
the waterfalls stage $n=n_p$. 

\subsection{$\bm \delta N$-formalism}
\label{sec:deltaN}

We now evaluate the curvature perturbation from inflation
by using the $\delta N$-formalism~\cite{Sasaki:1995aw}.
In this formalism, one first calculates the number of $e$-folds
for the background homogeneous universe from a fixed final 
epoch $t=t_f$ backward in time to an arbitrary initial epoch $t$,
$N(t\to t_f)$. The final epoch $t_f$ must be chosen such that at and
after $t=t_f$ the evolutionary trajectory of the universe is unique.

Then one considers the perturbation $\delta N$ by identifying
the final time slice to be a comoving (or uniform density)
hypersurface and the initial slice to be a flat hypersurface.
Naturally there will be fluctuations in the matter fields
on the initial flat slice which produce $\delta N$.
Then the resulting $\delta N$ is
equal to the comoving curvature perturbation on the
final hypersurface $\calR_c(\bmx,t_f)=\delta N(\bmx)$.
In the present model, we assume that the evolution of the
universe is unique, i.e. there is no isocurvature perturbation,
at and after the end of inflation. As before we use $n$
as the time instead of $t$ in what follows. Hence in particular,
$N=n-n_f$ where $n_f$ is the number of $e$-folds from
the critical point until the end of inflation.

There are a couple of points to be kept in mind when 
evaluating $\delta N$ in the present model. 
The first point is about the dependence of $N$ on the
fields $\phi$ and $\psi$. Since we have to relate the
fluctuations in these fields to $\delta N$, we have to
express $N$ in terms of $\phi$ and $\psi$. Namely, we
need an expression like $N=N(\phi(n),\psi(n))$.

Going backward in time from the end of inflation
up to the end of the waterfall stage $n_p<n<n_f$,
since the value of the waterfall field is determined by the 
value of the inflation, $\psi^2=\psi^2_{min}(\phi)$, 
with the identification $\psi=\sqrt{\langle\delta\psi^2\rangle}$,
$N$ is a function of only $\phi$, $N=N(\phi)$.
Since the scales that leave the horizon after the waterfall
stage $n>n_p$ are presumably too small to be of cosmological 
interest, we focus on the stage $n<n_p$ in the following discussion.

Here the very important point is that the time
when the waterfall stage ends, $n=n_p$, depends
heavily on the initial value of $\psi$ at the beginning of 
the waterfall stage, $\psi^2(n=0)$, hence on
the amplitude of the fluctuations $\langle\delta\psi^2(0)\rangle$.
This implies that $N$ is a function of both $\phi$ and $\psi^2$
through its dependence on $n_p$,
\begin{eqnarray}
N=N\Bigl(\phi(n),\psi^2(n)\Bigr)\quad \mbox{at}~n<n_p\,.
\end{eqnarray}

The second point is about the choice of the initial hypersurface.
In the present model the comoving curvature perturbation
due to $\delta\phi$ is conserved on superhorizon scales. Therefore, 
instead of choosing the time of horizon crossing for
each comoving scale, we can take the initial slice to be that
at the time of transition $n=0$ but with $\delta\phi$ 
given by that on the flat hypersurface at its horizon crossing as given by Eq.~(\ref{deltaNphi}) below. Therefore for $\phi$-dependence in $N$, we have
\begin{eqnarray}
\delta_\phi N\equiv
N(\phi_c+\delta\phi_c,\psi)-N(\phi_c,\psi)
=N_\phi(\phi+\delta\phi\to\phi_c)-N_\phi(\phi\to\phi_c)
\,,
\label{Nphi}
\end{eqnarray}
where $N_\phi(\phi\to\phi_c)$ is the number of $e$-folds 
from the time when the inflaton had a value $\phi$ until 
the time of transition when $\phi=\phi_c$, and $\delta\phi$ and
$\delta\phi_c$ are, respectively, the values of the field fluctuation
at horizon crossing and at the time of transition.
The contribution $\delta_\phi N$ may be approximated
to linear order in $\delta\phi$ as it is the same as the 
conventional slow-roll single-field contribution which
contains negligible non-Gaussianity. 
Noting
\begin{eqnarray}
\frac{\partial N}{\partial\phi}\delta\phi
=-\frac{H}{\dot\phi}\delta\phi=-\frac{\delta\phi}{\phi'}
=\frac{1}{r}\frac{\delta\phi}{\phi}\,,
\end{eqnarray}
where we have used Eq.~(\ref{phif}), we can evaluate $\delta_\phi N$
as
\begin{eqnarray}
\delta_\phi N=\frac{1}{r}\frac{\delta\phi_c({\bm x})}{\phi_c}
=\frac{1}{r}\frac{\delta\phi(n,{\bm x})}{\phi(n)}
=\frac{e^{rn}}{r}\frac{\delta\phi(n,{\bm x})}{\phi_c}\,,
\label{deltaNphi}
\end{eqnarray}
where we have used the fact that $\delta\phi$ on superhorizon
scales has the same time dependence as the background $\phi$.
When evaluating the spectrum, as usual, for a given scale $k$
we choose $n$ to be the time of horizon crossing $n=\ln(k/k_c)$ 
at which we have $\langle\delta\phi^2_k\rangle=H^2/(2\pi)^2$.

The contribution from $\delta\psi$ can be taken into
account by taking the difference between the number of $e$-folds
with $\delta\psi^2$ and $\psi^2=\langle\delta\psi^2\rangle$,
\begin{eqnarray}
\delta_\psi N=N(\phi_c,\delta\psi^2)-N(\phi_c,\langle\psi^2\rangle)\,,
\end{eqnarray}
where $\delta\psi^2$ and $\langle\delta\psi^2\rangle$
are to be evaluated at the time of transition $n=0$.
Rigorously speaking, for scales with $k>k_c$,
it should be evaluated at its horizon crossing $n=n_k(>0)$.
However, as noted at the end of the previous subsection,
we have conveniently included the modes $k>k_c$ also
on the hypersurface at $n=0$ in $\delta\psi^2(0,{\bm x})$
by formally extending their evolutionary behavior back to $n=0$
as if they were already outside the horizon. 
Hence one may calculate $\delta_\psi N({\bm x})$ by just
considering the fluctuations in $\delta\psi^2(0,{\bm x})$
without separating the modes into the two ranges 
$k<k_c$ and $k>k_c$. A justification of this point is
discussed in more detail in Appendix~\ref{betterapprox}.

Equation~(\ref{np-av}) shows how the number of $e$-folds
from the time of transition until the end of the waterfall stage
depends on the magnitude of the initial fluctuations 
$\langle\delta \psi^2(0)\rangle$.
This implies that the total number of $e$-folds of 
the waterfall stage fluctuates
from one horizon patch to another due to the fluctuations
in $\delta\psi^2(0,{\bm x})$, giving rise to the curvature
perturbation. Namely, we have
\begin{eqnarray}
\left(1+\frac{\delta n_{p}({\bm x})}{n_p}\right)^2
= 1 -\frac{1}{2 \kappa_\psi n_p^2}
\ln\left(\frac{\delta \psi^2(0,{\bm x})}
               {\langle \delta \psi^2(0) \rangle}\right)\,,
\label{deltanp}
\end{eqnarray}
where $\delta n_{p}({\bm x})$ is the spatial fluctuation 
in the number of $e$-folds from the time of transition until
the end of the waterfall stage. Apparently we have
$\delta_\psi N=\delta n_{p}({\bm x})$.

In the limit $\delta n_{p}/n_p\ll1$, we can linearize
the above to obtain
\ba
\label{deln-com}
\delta_\psi N=\delta n_{p}
= \frac{-1}{4 \kappa_\psi {n}_p} 
\ln \left(\frac{ \delta \psi ^2 (0,\bm{x})}
{ \langle \delta \psi^2(0) \rangle} \right)\,.
\ea
One may further expand the above logarithm,
 by using $\ln (1+x)=x - x^2/2+\cdots$, to find
\ba
\label{dn-p}
\delta_\psi N
= \frac{-1}{4\kappa_\psi n_p}
 \left[\frac{ \Delta \psi ^2 (0,\bm{x})}
{ \langle \delta \psi^2(0) \rangle}
 -\frac{1}{2} \left(\frac{\Delta \psi ^2 (0,\bm{x})}
{ \langle \delta \psi^2(0) \rangle}\right)^2 +\cdots\right]\,,
\ea
where, following Eq. (\ref{Delta-def}), we have
$
\Delta \psi ^2 (0,\bm{x}) 
= \delta \psi ^2 (0,\bm{x})- \langle \delta \psi^2(0) \rangle
$.

Here a comment is in order. One might worry that
there could be a contribution to $\delta n_{p}$
from the fluctuations in $\phi$ through the $\phi$-dependence
of $\psi_{min}^2$. Indeed there exists some contribution
because the value of $\phi$ at the end of the waterfall
stage depends on the number of $e$-folds from the critical
point as $\phi(n_p)\simeq\phi_ce^{-rn_p}$. However, one
can easily show that this contribution adds a term
$-r\delta n_{p}$ to the terms inside the square brackets in 
Eq.~(\ref{dn-p}), hence gives a negligible correction
for $\kappa_\psi  n_p/r\sim \beta\, n_p\gg1$.

\begin{figure}
\centerline{\includegraphics[scale=.4]{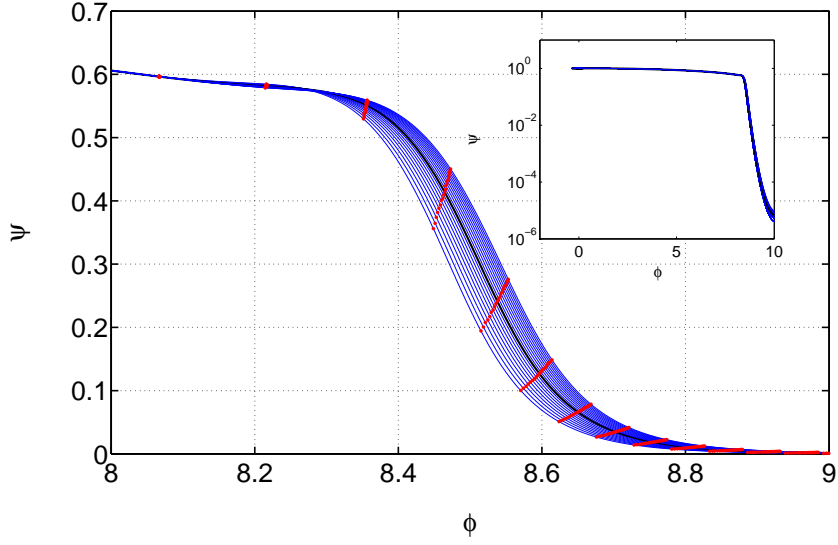}}
\caption{ Larger figure: Contour plot of constant 
$ N$ surfaces in the field-space are shown where $\psi$ and $\phi$
are in $M_P$ units.  Solid blue lines 
correspond to different patches of the Universe with different 
initial amplitude of the quantum fluctuation $\delta \psi^2 (0,\bm{x})$
 at the critical point. The solid black curve denotes 
the trajectory with $\delta \psi^2(0,\bm{x})= \langle \delta \psi^2 \rangle $. This figure shows the time interval between the 
critical point and few e-folds after the completion of the transition.
 Red contours show the constant $N$ contours. Counting the number of 
$e$-folds from the end of inflation, once the quantum back-reactions of  
$\delta \psi$ become important the constant $N$ contours incline towards 
the constant $\psi$ lines while during the earlier stage of inflation
the constant $N$ lines coincide with constant $\phi$ lines. 
Smaller (inner) figure: Trajectories of different patches of 
the Universe with different initial amplitudes of quantum fluctuations
 $\delta \psi^2(0,\bm{x})$ at the critical point. This 
figure provides an extended range for the interval between the critical 
point and the end of inflation.  
The numerical parameters are the same as in Fig.~1.
 }
\label{q-f-fr} 
\end{figure}

After the completion of the waterfall stage the trajectory in the
 two-dimensional field space $(\phi,\psi)$ is essentially unique,
with $\psi$ being completely subject to $\phi$, as can be seen from
Fig.~\ref{q-f-fr}, and the end of inflation
is synchronized on the uniform $\phi$ hypersurface $\phi=\phi_f$
as in the case of single-field slow-roll inflation.
This implies that there will be no contribution to the
total $\delta N$ from the era after the waterfall stage.

As we mentioned before, the field trajectory is unique
after the waterfall stage until the end of inflation.
Setting $n=n_f$ and $\phi^2(n_f)=\phi_f^2$ in Eq.~(\ref{phiend}),
the number of $e$-folds for this period $\Delta n_{ch} \equiv n_f-n_p$ can be
obtained as
\begin{eqnarray}
\label{n-chaotic}
n_f-n_p=\frac{\phi_p^2-\phi_f^2}{4M_{pl}^2}\,.
\end{eqnarray}
Hence the fluctuation in the number of $e$-folds
from this slow-roll stage, $\delta  (\Delta n_{ch})$ is given by
\begin{eqnarray}
\delta  (\Delta n_{ch})=\frac{\delta (\phi_p^2)}{4M_{pl}^2}
\simeq-r\,\frac{\phi_c^2}{2M_{pl}^2}\delta n_{p}\,.
\label{deltansr}
\end{eqnarray}
As is clear from this expression, $\delta  (\Delta n_{ch})$ is suppressed
by the factor $r$ ($\ll1$) relative to $\delta n_{p}$.

To summarize, the contribution to $\delta N$
after the transition is equal to $\delta_\psi N$ to a good approximation.
Hence the total $\delta N$ is expressed as
\begin{eqnarray}
\delta N=\delta_\phi N+\delta _\psi N\,,
\label{tot-delN}
\end{eqnarray}
where $\delta_\phi N$ is given by Eq.~(\ref{deltaNphi})
and $\delta_\psi N$ is given approximately by Eq.~(\ref{dn-p}).
We note that, if we express Eq.~(\ref{dn-p}) in the form
\begin{eqnarray}
\delta_\psi N=N_{,\psi^2 } \Delta\psi^2
+\frac{1}{2}N_{,\psi^2,\psi^2}(\Delta\psi^2)^2+\cdots,,
\label{dNexpand}
\end{eqnarray}
we have the relation among the coefficients,
\ba
\label{dN-dpsi}
N_{,\psi^2 } 
= - \langle \delta \psi^2 \rangle N_{,\psi^2,\psi^2}
 = \frac{-1}{4 \kappa_\psi{n}_p} 
\frac{1}{\langle \delta \psi^2 \rangle}\,.
\ea
This is a useful relation when one wants to evaluate
the non-Gaussianity in this model. 

\section{Power spectrum and Correlation functions}
\label{sec:corr}
This section is devoted to the calculations of the power spectrum
and correlation functions of the curvature perturbation. We find that the 
curvature perturbation induced from the waterfall is too large to be 
normalized to the COBE/WMAP normalization.

\subsection{Power Spectrum and spectral tilt}
The most important result of the $\delta N$ formalism is that one can 
identify the curvature perturbation on the final comoving surface
${\cal R}_c({\bm x})$, which is conserved thereafter,
by the fluctuation in the number of $e$-folds from an initial flat
surface to the final comoving surface, 
${\cal R}_c({\bm x})=\delta N({\bm x})$.
Taking the Fourier component of Eq.~(\ref{tot-delN}) with the help of
Eqs.~(\ref{deltaNphi}) and (\ref{dn-p}), to leading order we obtain
\ba
{\cal R}_{\bmk} = -\frac{1}{4 \kappa_\psi{n}_p}
 \frac{\Delta\psi^2_{\bmk}(0)}{\langle\delta\psi^2(0)\rangle}
  +\frac{1}{r}\left(\frac{k}{k_c}\right)^r
\frac{\delta \phi_{\bmk}(n_k)}{\phi_c}\,.
\ea
Using the fact that
 $\left(\Delta \psi^2 \right)_{\bmk}=\left(\delta \psi^2 \right)_{\bmk} $,
 the two point correlation function of the curvature perturbations 
can be read as
\ba
\label{cor-cur}
\langle {\cal R}_{\bmk} {\cal R}_{\bmk'} \rangle
  =\frac{1}{16 \kappa_\psi^2 {n}^2_p} 
\,\frac{\langle \left(\delta \psi^2 \right)_{\bmk}
  \left(\delta \psi^2 \right)_{\bmk'}\rangle_{0}}
  {\langle  \delta \psi^2\rangle_0^2} 
 +\frac{1}{r^2}\left(\frac{k}{k_c}\right)^{2r}
\frac{\langle\delta\phi_{\bmk}\delta\phi_{\bmk'}\rangle_{n_k}}{\phi_c^2}\,,
\ea
where the second term is the contribution of the original inflaton field
and the first term is the induced curvature perturbation from
the waterfall field. 


To calculate the power spectrum of the curvature perturbation we first
 start off by calculating the correlation function of $\delta \psi^2$ 
at the critical point, $n=0$. As shown in \cite{Gong:2010zf},
this correlation takes the form,
\ba
\label{corr-2}
\Big \langle \left( \delta \psi^2\right)_{\bmk}
 \,\left( \delta \psi^2\right)_{\bmk'}  \Big \rangle_0 
&\equiv&P_{\delta \psi^2 }(k)(2\pi)^3\delta^3 (\bmk+\bmk')
\nonumber\\
&=&2 \int d^3 q \vert \delta \psi_q(0) \vert^2 \,\vert 
\delta \psi_{\vert \bm{k}-\bm{q}\vert }(0) \vert^2 
\, \delta^3 (\bmk+\bmk')\,.
\ea

Here we estimate this integral which would suffice for our discussion here.
In Appendix \ref{curvature-rigorous} we provide a more rigorous treatment
of calculating this integral. 
Let us  first estimate the contribution of the infrared modes in this 
integral, say $|q| \lesssim |k|$. The contribution of the long wavelength
 modes to the two point correlation function approximately can be found to be
\ba
P_{\delta \psi^2}^{IR} 
= \dfrac{2\vert \delta \psi_{k }(0) \vert^2}{(2\pi)^3} 
\int \limits_{|q| \lesssim |k|} d^3 q \vert \delta \psi_q(0) \vert^2 \,.
\ea
The above integral counts the effects of large modes in the 
convolution. Note that we take the term 
$\vert \delta \psi_{\vert \bm{k}-\bm{q}\vert }(0) \vert^2$ out of 
the integral which is possible for small $q$.  To simplify the analysis,
 one can  take the upper bound to be $k_c$ so
the contribution of the infra red momenta in the convolution can be read as
\ba
\label{tpcf}
P_{\delta \psi^2}^{IR}
 = \vert \delta \psi_{k }(0) \vert^2\, \dfrac{H^2}{4 \pi^2}
 \,\int_{-\infty}^{0} dn_q e^{-2 \kappa_\psi n_q^2}\, 
\, = \dfrac{1}{2} \langle \delta \psi^2 (0) \rangle 
\vert \delta \psi_{k }(0) \vert^2\,.
\ea

Now we estimate the contribution of the small modes in the 
convolution of integral in Eq.~(\ref{corr-2}).  By the small modes 
we mean the modes for which, $ |\mathbf{q}| >  |\mathbf{k}|$. 
By this assumption, one can simply find
\ba
P_{\delta \psi^2}^{UV}
 = \dfrac{1}{(2\pi)^3}\int \limits_{|q| \gtrsim |k|} d^3 q \,
 \vert \delta \psi_q(0) \vert^4 \, 
\, &\simeq& \dfrac{H^4}{16 \pi^2 k_c^{3}}
 \int_{\ln k/k_c}^{\infty} d n_q e^{-3 n_q - 4 \kappa_\psi n_q^2} \, .
\ea
As we are interested in the $|q| \gtrsim |k|$ to lower bound 
simply taken to be $k=q$ in r.h.s. The second order term in the exponent
 suggests a natural cut-off $n_q < 1/\sqrt{2 \kappa_\psi}$ 
for the integral so one can simply ignore the quadratic term of $n_q$ 
in the exponent to obtain
\ba
\label{uv-contribution}
P_{\delta \psi^2}^{UV} \sim \, \dfrac{\sqrt{\kappa_\psi}}{2 } 
\langle \delta \psi^2(0) \rangle \vert \delta \psi_k(0) \vert^2 \, .
\ea
One readily finds that the ultraviolet contribution from 
Eq.~(\ref{uv-contribution}) is suppressed compared to the infrared 
contribution, Eq.~(\ref{tpcf}),  by a factor $\sqrt{\kappa_\psi} \ll 1$ 
so the power spectrum  from the $\psi$ contribution mainly comes from
the infrared modes and $P_{\delta \psi^2} \simeq P_{\delta \psi^2}^{IR}$.

Finally from Eq.~(\ref{cor-cur}) the curvature perturbation spectrum 
per unit logarithmic frequency interval,
\ba
{\cal P_{R}} \equiv \frac{k^3}{2 \pi^2} P_{\cal R}
 = \frac{k^3}{2 \pi^2} |{\cal R}_{k}|^2
\ea
is obtained to be 
\ba
\label{curv-final}
{\cal P_{R}} &\simeq& {\cal P_R}^{IR} + {\cal P_R}^{(\phi)} \nonumber\\
 &=&\left( \dfrac{1}{4 \kappa_\psi n_p}\right)^2
 \frac{k^3}{4 \pi^2}
\dfrac{\vert \delta \psi_k(0)\vert^2}{\langle \delta \psi^2(0) \rangle}
 + \frac{g^2}{4 \pi^2 r^2 \beta} e^{2 r n_k}
 \nonumber\\
 &\simeq&  \left( \dfrac{1}{4 \kappa_\psi n_p}\right)^2 
\sqrt{\dfrac{\kappa_\psi}{2 \pi}} e^{-2 \kappa_\psi n_k^2}
  + \frac{g^2}{4 \pi^2 r^2 \beta} e^{2 r n_k}\,.
\ea
We need $g\ll1$ so the contribution from the inflaton field,
 ${\cal P^{(\phi)}_{R}}$ is sub-leading compared to ${\cal P_R}^{IR}$. 

With the curvature perturbation power spectrum dominated by ${\cal P_R}^{IR}$
the spectral index of the curvature perturbation can be easily found as
\ba
\label{ns}
n_s -1 = \frac{d \ln({\cal P_R} )}{d \ln k}
 = - 4 \kappa_\psi n_k
=-4\kappa_\psi \ln(k/k_c)\,.
\ea
Since $\kappa_\psi  \ll 1$, the curvature perturbation is
 nearly scale invariant but it has a slight running.  It runs
from blue for modes which exit the horizon before the 
transition to red for modes which exit the horizon after 
the transition. One can easily quantify this running as
\ba
\frac{d n_s}{d \ln k} = -4 \kappa_\psi\,.
\ea
This should be compared with the standard hybrid inflation 
where it is found that $n_s \simeq 4$~\cite{Lyth:2010ch, Abolhasani:2010kr,
 Fonseca:2010nk, Gong:2010zf, Lyth:2010zq}, a very blue-tilted spectrum.
This is the key property which renders  the  curvature perturbations
 from the waterfall field harmless in standard hybrid inflation. 

Now we face the fatal problem in our model. We see from 
Eq.~(\ref{curv-final}) that the curvature perturbation power spectrum 
cannot be normalized to the required COBE/WMAP normalization
${\cal P_{R}} \sim  2\times 10^{-9}$. With $\kappa_\psi  \ll 1$ 
and $4 \kappa_\psi  n_p \sim 1$ one finds that ${\cal P_R}^{IR}$ 
is many orders of magnitude bigger than the COBE amplitude. 
In comparison, in standard hybrid inflation this problem does not show up. 
As shown in recent analysis \cite{Lyth:2010ch, Abolhasani:2010kr,
 Fonseca:2010nk, Gong:2010zf, Lyth:2010zq} the curvature perturbations 
from the waterfall field in standard hybrid inflation with
 $\phi_c \ll M_P$ and $\kappa_\psi  \gg 1$ scales like 
$({k}/{k_c})^3 \sim e^{-3N_c}$ which is totally negligible
 on the cosmological scales. 

One may think under what conditions the curvature perturbations induced
from the waterfall can be kept sub-dominant on the cosmological scales.  To address this let us first estimate the minimum e-foldings required  between the transition time and the time of horizon crossing of the large scale (CMB) modes. We would like to
impose the following constraint on the amplitude of these modes
\begin{eqnarray} 
{\cal P_R}^{IR}(k^{CMB}) < 2 \times 10^{-9}\, ,
\end{eqnarray}
in which $k^{CMB}$ is the momentum of cosmological sizes observed on CMB.  
By using  Eq. (5.9),  for typical values of our parameters, this yields
\begin{eqnarray}
\label{abs-nk}
|n_{k^{CMB}}| > 23 \times \sqrt{\dfrac{1}{2 \kappa_{\psi}}}  \, .
\end{eqnarray}
To have $50-60$ $e$-folds after the transition, we must have $n_p+\Delta n_{ch}=50-60$. 
But if $\Delta n_{ch}$ dominates it is essentially equivalent to a chaotic 
inflation model which is not interesting. Hence let us assume $n_p\gtrsim50$ or so. 
Then Eq.~(\ref{abs-nk}) above implies that this is possible only for 
$n_{k}^{ CMB}<0$. But then the number of $e$-folds form the CMB scale till 
the end of inflation becomes too large. Hence it is impossible to hide the 
curvature perturbation from the waterfall field for the range of the 
parameters we consider unless the final chaotic inflation stage dominates.

Now comes the natural question that what is the source of the
 problem? The key to this questions is the lack of a classical 
background $\psi$ trajectory so the quantum fluctuations 
$\delta \psi_k(n)$ determine the effective classical trajectory 
$\sqrt {\langle \delta \psi^2(n) \rangle}$. The ratio of quantum 
fluctuation $\delta \psi_k$ to this effective background trajectory 
represents the amplitude of the quantum fluctuations. 
This ratio in the coordinate space is 
\ba
\dfrac{k^3}{(2 \pi)^3}
\dfrac{\vert \delta \psi_k(0) \vert^2}{\langle \delta \psi^2(0) \rangle}
 \sim \sqrt{\kappa_\psi} \, ,
\ea
which is not small enough to be normalized to the observations. 
To see this better  compare the situation with the standard 
chaotic models. In models of chaotic inflation, the  fluctuations 
are given by $\delta \phi / \phi$ in which $\delta \phi \sim H$ 
while the value of the background $\phi$ field is another free 
parameter which can be tuned independently to get the right normalization. 
In other words, in chaotic models  there is a well-defined 
background classical $\phi$ field which exists independent of the
quantum fluctuations $\delta \phi$. However, in our model there is 
no classical background  $\psi$ field. Instead the effective background 
waterfall field $\sqrt{\langle \delta \psi^2(n) \rangle}$ is constructed 
from the quantum fluctuations $\delta \psi_k(n)$ and as a result there 
is no free parameter to tune the relative ratio 
$k^3{\vert\delta\psi_k(0)\vert^2}/{\langle \delta \psi^2(0) \rangle}$
to a very small value.

Having this said, one can naturally ask why this problem does not appear
in standard hybrid inflation model. As mentioned above, the main feature
 in standard hybrid inflation is that the large scale waterfall 
perturbations are highly blue-tilted, scaling like $({k}/{k_c})^3$.
On the other hand, the main contribution into 
$\sqrt{\langle \delta \psi^2 \rangle}$ comes from small scales, modes
 which become tachyonic but remained sub-horizon till end of inflation. 
As a result the cosmological large scales, corresponding to the current 
Hubble size,  are suppressed compared to these small scales by the ratio 
$ ({k}/{k_c})^3 \sim  e^{-3N_c/2}$ and therefore are completely 
negligible in the curvature power spectrum. In contrast, in our model 
the power spectrum is nearly scale invariant, as can be seen from 
Eq.~(\ref{ns}), so the large scale modes are not significantly
suppressed compared to the small scales and both of these modes 
contribute equally into $\sqrt{\langle \delta \psi^2 \rangle}$.
As a result, one cannot tune the ratio 
$k^3{\vert \delta \psi_k(0) \vert^2}/{\langle \delta \psi^2(0) \rangle}$
to a very small value. 

In passing we note that in hybrid inflation with a sharp phase 
transition if one reduces $N_c$ so the phase transition happens 
early during inflation, i.e. $N_c  \sim 5$, then the large scale 
modes may survive the suppression.
 We address this case in the next section.

\section{General parameter space of hybrid inflation}
\label{general-limit}

In last three sections we have focused on the model where 
$\phi_c^2 > 12M_P^2   $ and $\kappa_\psi  \ll1$ so inflation has 
three extended stages where $N_{tot} = N_c + n_p+ \Delta n_{ch}$.
We have particularly studied this limit of the parameter space in great 
details for two reasons. First, this limit was not studied previously 
in the literature, namely the effects of the waterfall phase transition on 
curvature perturbations. Secondly, this limit of parameter space 
provides us a specific example in which one finds an unexpected result
 that the contribution of the waterfall field to the power spectrum  
is too big to satisfy the COBE normalization. As we have seen in 
the previous section, this is because the effective background 
trajectory $\langle \delta \psi^2 \rangle$ gets more or less equal 
contributions from both the large scale and small scale  quantum 
fluctuations $\delta \psi_k$ and as a result  there is no free parameter
 to suppress the ratio 
${\vert \delta \psi_k(0) \vert^2}/{\langle \delta \psi^2(0) \rangle}$.

For the other limit of parameter space, we have standard hybrid 
inflation with $\phi_c <M_P$ and $\kappa_\psi >1$ so inflation ends 
quickly after the phase transition. As shown in~\cite{Lyth:2010ch, 
Abolhasani:2010kr, Fonseca:2010nk, Gong:2010zf, Lyth:2010zq} the 
induced curvature perturbation from the waterfall field is highly 
blue-tilted and is completely negligible on large scales. 
 We would like to search the parameter space of hybrid inflation 
in general and examine the dynamics of waterfall phase transition and 
its contribution to power spectrum. We would like to see under what 
conditions  the results vary  between the results of standard hybrid inflation 
and the results we found in the previous sections. 

As mentioned in section \ref{sec:model}, two major classes of the 
parameter space were determined by the ratio $\phi_c^2 / 12M_P^2$.
In the previous sections we have studied the case $\phi_c^2 > 12M_P^2 $ 
and $\kappa_\psi  \ll 1$ in details. Now it is time to consider other
limits of the parameter space. 

\subsection{ $\phi_c^2 > 12M_P^2 \quad ,  \quad    \kappa_\psi\gg 1$}

Unlike the case studied in the previous sections, the condition 
$\kappa_\psi\gg 1$ results in a sharp phase transition  similar 
to standard hybrid inflation. However, in contrast to standard 
hybrid inflation in which inflation ends shortly after the waterfall,
 in this case with $\phi_c^2 > 12M_P^2$ inflation continues for 
an extended period after the transition.
 On the other hand, knowing that $\alpha \sim 1/N_{tot} \ll1$, to have
$ \kappa_\psi\gg 1$ one requires that $\beta \gg N_{tot}$, 
i.e. $\beta \sim 10^2-10^3$.
As in previous sections, inflation has three stages and 
$N_{total} = N_c + n_p + \Delta n_{ch}$.  Here we first briefly 
review the background dynamics and then compute the final 
curvature perturbations using the $\delta N$ formalism.
\begin{enumerate}
\item[Stage.1]
The first stage corresponds to time before phase transition, 
i.e. $N<N_c$ when the inflaton field reaches to the critical value,
 $\phi = \phi_c$, after which the waterfall field becomes tachyonic. 
Fluctuations of the inflaton field, $\delta \phi$, which exit the
 horizon before the critical point cause variations on the duration
 of this period, $\delta N_c$. By using Eq. (\ref{phif}),
 one has $\phi_c = \phi_i e^{-r n_i}$, in which  $\phi_i$ and $n_i$
 respectively are the value of the inflaton field and the time of 
horizon crossing for a specific mode. So the variation of $N_c$
 can be read as
\ba
\label{delnc}
\delta N_c = \dfrac{-1}{r} \left( \dfrac{\delta \phi}{\phi} \right) _{k<k_c}
\ea
Since the next stage of inflation starts with $\phi=\phi_c$ for each 
patch therefore the information associated to these perturbations 
is encoded in $\delta N_c$

\item[Stage.2]
This stage is defined as the period after the transition until the 
completion of transition, $n_p$, 
when $\psi$ becomes 
\ba
\langle\delta\psi^2\rangle=\psi_{min}^2(\phi) \equiv
 \frac{M^2}{\lambda} -\frac{g^2}{\lambda}\phi^2\,. 
\ea
As in standard hybrid inflation this period is very short, 
$n_p \lesssim 1$, but since the inflaton field is super-Planckian 
inflation can resume for the third stage in the form of chaotic 
inflation. As  shown 
in~\cite{Abolhasani:2010kr,Lyth:2010zq,Gong:2010zf,Fonseca:2010nk} 
the spectrum of the waterfall  fluctuations which become tachyonic
shortly after the transition is highly  blue-tilted while for 
modes which become tachyonic later on the power spectrum is highly 
red-tilted. For modes in between the spectral index is running 
significantly from blue to red so one should avoid these modes in 
contributing to the total power spectrum. 
As we discussed before the amplitude  of cosmological large 
scales is suppressed by the factor
$e^{-3 N_c/2}$ compared to the small scales. Since the small scale 
modes have the dominant contributions into 
$\sqrt{\langle \delta \psi^2 \rangle}$ as the effective classical
 trajectory, the contribution of the large scales is suppress in
 the power spectrum by the ratio $e^{-3 N_c}$.
Below we will provide a bound on $N_c$ in which one can safely neglect 
the contribution of large scale curvature perturbations.

\item[Stage.3]

This stage is the period of chaotic inflation after the transition, $n>n_p$.
The end of this period is defined by an explicit value of the 
$\phi$ field. The $\psi$ field in this period does not play any role
 because it is held on its minimum. The inflaton fluctuations which 
exit the horizon after critical point can change the number of $e$-folds
 during chaotic period.  From  Eq.~(\ref{n-chaotic})  one has
\begin{eqnarray}
n_f-n_p=\frac{\phi_p^2-\phi_f^2}{4M_{pl}^2}\,.
\end{eqnarray}
\end{enumerate}

To calculate curvature perturbations using $\delta N$ formalism first
 concentrate on the fluctuations which exit the horizon before the 
critical point. One can borrow  the results from Section~\ref{sec:deltaN}.
 The only difference is the dynamics of the waterfall field. 
In general suppose that the waterfall field has the following 
form after the transition 
\ba
\psi ^2 (n) = \psi^2(0) \, e^{2 \, f(n)}.
\label{tachyonicgrowth}
\ea
For our case at hand, in which the phase transition happens
 similar to standard hybrid inflation~\cite{Abolhasani:2010kr,Gong:2010zf},
we have
\begin{eqnarray}
f(n)=\frac{2}{3}\sqrt{6 \kappa_\psi}n^{3/2}\,.
\label{fnform}
\end{eqnarray}
The time at the end of transition can be found by using the
equation,
\ba
\psi_{min}^2 = \delta \psi^2(0) \, e^{2 \, f(n_p)} \, .
\ea
As before, ignoring the slight dependence of $\psi_{min}(\phi)$ on $\phi$, 
the variations of the  phase transition duration to the first order
 can be read as 
\ba
\label{deln-psi}
\delta \, n_p (\mathbf{x}) \simeq \dfrac{-1}{2 \, f'(n_p)} \,
 \dfrac{\Delta \psi^2(\mathbf{0,x})}{\langle \delta \psi^2(0)\rangle},
\ea
in which 
$\Delta \psi^2(0,\bmx)=\delta\psi^2(0,\bmx)-\langle\delta\psi^2(0)\rangle$. 

Adding up the contributions from the inflaton field, Eq.~(\ref{delnc}),
 and the waterfall field, Eq.~(\ref{deln-psi}),
one finds the final curvature perturbation for  modes $k<k_c$,
\ba
{\cal R}_c (k)= \delta N = \dfrac{-1}{2 \, f'(n_p)} \, 
\dfrac{\left(\delta \psi^2\right)_k}{\langle \delta \psi^2(0)\rangle}
 +\left(\dfrac{\delta\, \phi}{r\,\phi}\right)_k\,;  \quad \quad k<k_c\,.
\ea

Finding the curvature perturbation for modes which leave the horizon 
after waterfall, $k>k_c$, requires some careful considerations. 
We relegate the details into Appendix \ref{app:deln-k>kc} and quote
 the final result,
\begin{eqnarray}
{\cal R}_c (k)= \delta N
=-\frac{1}{2f'(n_p)}
\frac{\left(\delta\psi^2 \right)_k}{\langle \delta \psi^2(0) \rangle}
+\left[1-\frac{f'(n_k)}{f'(n_p)}\right]
 \left(\dfrac{\delta\phi}{r\phi}\right)_k\,; \quad \quad k>k_c\,.
\end{eqnarray}
As is seen from this result, the curvature perturbation spectrum is 
smoothly taken over from $\delta\phi$ to $\delta\psi^2$ as  
$n$ increases from $0$ to $n_p$.

As mentioned before, during the second stage the power spectrum 
can run significantly from blue to red. To avoid  this one has to
 make sure the phase transition happens late enough during inflation
 so the contributions of these modes are negligible on large scale
 power spectrum.
Following the analysis in \cite{Gong:2010zf} (for a method 
different than $\delta N$ see also \cite{Abolhasani:2010kr}) 
the contribution of the waterfall field in total curvature perturbation 
is \footnote{Note that the notation $\alpha$ used in \cite{Gong:2010zf} 
can be translated to our $\kappa_\psi$ here as 
$\alpha =\sqrt{6 \kappa_\psi}$} 
\ba
{\cal P_R}^{(\delta \psi)} \simeq 
\dfrac{\kappa_\psi^{-11/3}}{n_p} \left( \dfrac{k}{k_c}\right)^3 \,.
\ea
The cosmological large scales relevant for CMB exit the horizon 
about $60-53$ $e$-folds before the end of inflation, so one
 should check that ${\cal P_R}^{(\delta \psi)}$ is  smaller than 
the observed COBE normalization ${\cal P_R} \sim10^{-10}$. This yields
\ba
\kappa_\psi^{-11/3} e^{-3(N_c-7)} \ll 10^{-10}\,,
\ea
which leads to the following constraint
\ba
\label{Nc-ineq}
N_c > 14.7 -\frac{11}{9} \ln \kappa_\psi \, .
\ea
The above relation indicates that if one takes $\kappa_\psi$ large 
enough then ${\cal P_R}^{(\delta \psi)}$ is so suppressed that the
 transition can take place for $N_c \gtrsim 1$.
 Below we study each stage in brief.

\subsection{$\phi_c^2 < 12M_P^2 $}

As discussed in section \ref{sec:model} inflation has four stages 
$N_{tot}= N_c + n_{trig} + \Delta n_{\phi-trans}+ \Delta n_{\psi-trans}$
in which the $\phi$ transition period is very short. 

\begin{enumerate}
\item[Stage.1]
This stage, corresponding to  $N<N_c$, is completely similar
 to the case studied before.

\item[Stage.2]
This stage corresponds to the case where the back-reaction of 
$\psi$ on $\phi$ becomes important triggering it to move towards its
 global minimum $\phi=0$. The value of the $\psi$ at the time of 
$n_{trig}$ has no dependence on inflaton field so fluctuations of the 
inflaton field does not affect the duration of this period. 
Furthermore, one should now replace $n_p$ in previous subsection by 
$n_{trig}$ because the transition of $\psi$ shuts off at $n_{trig}$
 instead of $n_p$.  One can show that 
\ba 
\label{n-tri}
n_{trig}^2 = \dfrac{1}{2 \kappa_\psi} \,
 \ln  \left(\dfrac{\psi_{min}}{\psi(0)} \dfrac{\phi_c}{M_p} \right) \, .
\ea
As in the previous sub-section suppose after the phase transition 
$\psi \sim \psi_c e^{f(n)}$ where the form of $f(n)$ depends on the 
sharpness of the phase transition. Following the same steps as in 
Section~\ref{sec:deltaN}, the contributions of $\psi$  fluctuations
 into the variation of the waterfall transition number of $e$-folds is
\ba
\label{deln-2}
\delta n_{trig}
= \frac{-1}{2f'(n_{tri})} 
\ln \left(\frac{ \delta \psi ^2 (0,\bm{x})}
{ \langle \delta \psi^2(0) \rangle} \right)\,.
\ea
Regardless of the details of the transition, one deduces that 
the contribution of the waterfall field 
quantum fluctuations into the total curvature perturbations is 
\ba
{\cal R}_c (\delta \phi =0 ) = \delta n_{trig}
= \frac{-1}{2f'(n_{tri})} \frac{ \left( \delta \psi ^2 \right)_k}
{ \langle \delta \psi^2(0) \rangle} \,.
\ea

\item[Stage.3]

This stage corresponds to the short interval $\Delta n_{\phi-trans}$ 
when $\phi$  rolls quickly to its minimum.  One can still consider the 
vacuum domination in this period as the waterfall field is far from
 its global minimum. The equation of the inflaton field can be read
 as follows
\ba
\label{phi-after-tri}
\phi''+3\phi'+ \omega^2(n) \phi =0
\ea 
in which 
\ba
\omega^2(n) = \alpha + \dfrac{g^2}{H^2} \psi^2 
\ea
and $ \psi \sim \psi_{trig} e^{f(n)}$. Defining $\phi = e^{-3/2 n} \Phi$ one has
\ba
\label{phi-after-tri}
\Phi''+ \left(\dfrac{-9}{4}+\omega^2(n) \right)\Phi =0 \, .
\ea
Shortly after the triggering time the term containing $\omega^2$ becomes
 large and completely dominates over  $-9/4$ so one can use the 
WKB approximation to find the time dependence of the inflaton field 
during the fast roll towards $\phi=0$,
\ba
\phi \sim e^{-3n/2 } e^{\pm i\int ^{n} \omega(n') \md n'}.
\ea
This indicates $\phi$ scales like $e^{-3n/2}$ during its fast roll 
towards its minimum. The transition of $\phi$, triggered due to 
back-reaction of the waterfall field, can be considered to be complete 
when the interaction term $g^2 \phi^2 \psi^2$ becomes negligible compared
 to the bare mass of the waterfall field 
\ba
\phi^2 \ll \phi_c^2 \, .
\ea
Since the inflaton field slow rolls before the triggering point, 
its amplitude there is about $\phi_c$. One can readily concludes 
that the fast roll of inflaton field completes in 
\ba
\Delta n_{\phi-trans} \sim 1\,.
\ea
So one can consider that this transition is somewhat instantaneous. 

\item[Stage.4]
Soon after  $\phi$ settles to its global minimum $\phi=0$ the waterfall 
field experiences the large tachyonic mass $-M^2$. One can check  
that in this period $\psi$ starts growing 
like $\psi (n) \sim e^{\sqrt \beta n}$ towards its global minimum.
\end{enumerate}

Again, to find the total curvature perturbation one should add up 
different contributions to $\delta N$ from different periods 
specified above. If $\kappa_\psi  \gg 1$, corresponding to a 
sharp phaser transition, the situation is the same as in standard 
hybrid inflation. Since the inflaton field is sub-Planckian then 
inflation can not continue after the waterfall in the form of chaotic
 inflation so one basically requires $N_c \simeq N_t \sim 60$. 
Furthermore, the large scale curvature perturbations do not get
 contribution from the waterfall field fluctuations which are 
suppressed by $(\frac{k}{k_c})^3 \sim e^{-3N_c} \ll 1$.
 Note that with the assumption $\kappa_\psi  \gg 1$ we have 
$\beta \gg N_{total}$ i.e. $\beta \sim 10^2 - 10^3$.
 
 However, if we assume $\kappa_\psi  \ll 1$ so the phase transition 
is mild, then we face the same problem as discussed in 
section~\ref{sec:corr}, that is the curvature perturbations cannot
 be normalized to the COBE normalization. This can be seen from 
Eqs.~(\ref{deln-2})
which has the same structure as Eqs.~(\ref{deln-com}) and (\ref{dn-p}).

 \section{Summary and Discussions}
 In this paper we have looked into the parameter space of hybrid 
inflation. We have retained two key properties of the standard hybrid 
inflation scenario, the vacuum domination condition and the condition
 that the waterfall field is very heavy during early stage of inflation.
 The latter condition is equivalent to $\beta \gg 1$ causing the waterfall
 field to quickly rolls to its instantaneous minimum $\psi=0$. Besides 
these two conditions, we have relaxed other conditions on the hybrid 
inflation parameters and searched for the limits of the parameter space 
where one can obtain curvature perturbations with the right amplitude. 
As we have seen, two key parameters in our classification are  
$\phi_c/M_P$ and  $\kappa_\psi $ where the latter 
parametrizes the sharpness of the phase transition. For a sharp phase 
transition, as in standard hybrid inflation, one has $\kappa_\psi  \gg 1$, 
whereas for a mild phase transition $\kappa_\psi  \ll 1$. 
 
 We have shown that models with $\phi_c^2 > 12M_P^2 $ and
 $\kappa_\psi  \ll 1$ do not work because they can not satisfy the
 observed COBE normalization. Note that to have $\kappa_\psi  \ll 1$ 
while keeping the waterfall field heavy requires that $\beta \sim 10$ 
or so. In this model inflation has three extended stages, the first stage 
is before the waterfall, the second stage starts when the waterfall rolls
 mildly to its minimum while inflaton is slowly rolling. The final stage
 is similar to conventional chaotic inflation models. As we have seen, 
the dominant contributions into the curvature perturbation power spectrum 
come from the waterfall fluctuations. On the other hand, there is 
no background classical $\psi $ trajectory so one identifies 
$\sqrt{\langle \delta \psi^2(n) \rangle}$ as the effective background 
trajectory.   As a result, one finds that there is not a very small 
free parameter to tune the ratio 
$({k^3}/{(2 \pi)^3}){\vert\delta\psi_k(0)\vert^2}/{\langle\delta\psi^2(0)\rangle}\sim \sqrt{\kappa_\psi}$
 to the small observed value.
 This is in contrast with, say the $m^2 \phi^2$ chaotic model, 
in which $\delta \phi \sim H$ while $\phi$ is kept as an initial 
value free parameter so one can tune $\delta \phi/\phi$ into an 
arbitrary small number. On the other hand models with 
$\phi_c^2 > 12M_P^2 $ and $\kappa_\psi  \gg 1$ can be allowed. 
The contributions of waterfall field fluctuations are the same as 
in standard hybrid inflation. However, one has to satisfy 
inequality~(\ref{Nc-ineq}) in order to avoid a significant 
running of spectrum from blue to red on the CMB scales.

Models with $\phi_c^2 < 12M_P^2 $ and $\kappa_\psi  \gg 1$ are similar
 to standard hybrid inflation and one can satisfy the COBE/WMAP normalization.
 In these models inflation end abruptly after the waterfall phase
 transition. For having $\kappa_\psi  \gg 1$, one requires that 
$\beta \sim 10^2-10^3$. However, in models with  $\phi_c^2 < 12M_P^2 $ 
and $\kappa_\psi  \ll 1$ one encounters the same problem as above that 
the COBE/WMAP normalization cannot be satisfied.
This corresponds to $\beta \sim 10$ or so.

After these discussions, one concludes that models of hybrid 
inflation with the assumptions of vacuum domination and a mild 
phase transition do not work  because one can not normalize the 
curvature perturbation power spectrum to the required COBE/WMAP
normalization. This corresponds to models with $\kappa_\psi  \ll 1$
 and  $\beta \sim 10$ or so.  

\section*{Acknowledgement}

We would like to thank David Lyth, Mohammad Hossein Namjoo and David Wands for useful discussions.
 A.A.A. would like to thank YITP for the hospitality where this work started.
 H.F. would like to thank ICG for the hospitality where this work was completed.
M.S. would like to thank KIAS for the hospitality where the final part of
this work was done.
This work was supported in part by 
MEXT Grant-in-Aid for the global COE program at Kyoto University,
 "The Next Generation of Physics, Spun from Universality and Emergence,"
and by Korea Institute for Advanced Study under the KIAS Scholar program.
This work was also supported in part by JSPS Grant-in-Aid for Scientific 
Research (A) No.~21244033, and by JSPS Grant-in-Aid for Creative Scientific 
Research No.~19GS0219.

\appendix

\section{Justifying integrating out sub-horizon modes in Eq. (\ref{del-psi0-def})}
\label{betterapprox}

Here we justify our approximation in integrating the amplitudes of all of quantum modes regardless of being super-horizon or sub-horizon at critical point, given
by Eq. (\ref{del-psi0-def}). In order to employ the $\delta N$ formalism we have to start with
a well-defined background classical trajectory \cite{Wands:2000dp}.  In our model, there is no well-defined background trajectory for the $\psi$ field. However,  the accumulative effects of quantum modes $\delta \psi $ defines the effective $\psi $ background as 
$\sqrt{\langle \delta \psi^2  \rangle}$.  To find this average we should consider the modes which are at any time super-horizon. In other words, we renormalize the two point function with cut-off defined by the comoving momentum $\Lambda=k/a =H$. By the above discussion the background $\psi_b$ at any time can be read as follows
\ba
\psi^2_{b} = \langle \delta \psi^2  \rangle_{ren.} (n) = \int _{0}^{k=a H_0} \frac{d ^3 k}{(2 \pi)^3}  \vert \delta  \psi_k ^{(0)} \vert ~ e^{2 \kappa_\psi n^2} \, .
\ea
By using  Eq. (\ref{dpsibetter}) one can simplify the above integral
\ba
\psi^2_{b}  = \frac{H_0^2}{4 \pi^2} \int _{- \infty}^{n} d n_{ k} e^{-2 \kappa_\psi n^2_{ k }}  ~ e^{2 \kappa_\psi n^2}
\ea
In the above integral, at any given time time, only  the super-horizon modes are considered. The above integral can be find in terms of error-functions 
\ba
\psi^2_{b}  = \frac{H_0^2}{8 \pi^2} \sqrt{\frac{\pi}{2 \kappa_\psi}} \, \left(1+ \mathrm{Erf} \left( \sqrt{2 \kappa_\psi }n\right) \right) ~ e^{2 \kappa_\psi n^2} \,   .
\ea
By using the properties of the error-functions, one can verify that for the time 
$ n > 1/\sqrt{2\kappa_\psi} $ the above expression approaches quickly to 
\ba
\label{eff-back-s}
\psi^2_{b}  \simeq \frac{H_0^2}{4 \pi^2} \sqrt{\frac{\pi}{2 \kappa_\psi}} ~ e^{2 \kappa_\psi n^2} = \langle\delta \psi^2 (0) \rangle  ~ e^{2 \kappa_\psi n^2} 
\quad \quad  ( n > 1/\sqrt{2\kappa_\psi} )\, .
\ea
Now let us compare $1/\sqrt{\kappa_\psi}$ with $n_p$: the bigger is the value of $\sqrt{\kappa_\psi} n_p >1 $ the better gets our approximation Eq. (\ref{eff-back-s}).
Let us  now have an estimate for $\sqrt{\kappa_\psi} n_p$. From Eq. (\ref{np-av}) we have
\ba
\kappa_\psi n_p^2  \simeq -\frac{1}{2}  \ln \left( \frac{\beta}{\lambda} \sqrt{32 \pi^3 \kappa_\psi } \right) \, .
\ea
The term inside the square root is basically at the order of unity, therefore 
$\kappa_\psi n_p^2 \simeq -\ln \sqrt \lambda$. For $\lambda$ sufficiently small
one can arrange for $\kappa_\psi n_p^2 \gg 1$. For example in our numerical analysis, $\lambda \sim 10^{-9}$ which gives $\kappa_\psi n_p^2 \simeq 21$ so our approximation Eq. (\ref{eff-back-s}) is very good. Therefore for times well after the phase transition but before 
the phase transition completion one can use the approximation (\ref{eff-back-s}) as a good estimate for the background  $\psi$ field.

In the $\delta N$ formalism the evolution of curvature perturbations is considered only for super-horizon modes  \cite{Sasaki:1995aw, Wands:2000dp}. Working on super-horizon scales means smoothing out the sub-horizon perturbations in a patch. At any time super-horizon modes are the modes with the comoving wavelength larger that the Hubble radius, $k < a H$. Therefore one can integrate the curvature perturbations by considering only the super-horizon modes at a given time. For example in our model, at the time of phase transition there are many modes which are sub-horizon, but as time goes by these modes become super-horizon and should be considered in the background evolution. This point is what we considered in evaluating the average trajectory of $\psi$ field. Therefore to calculate $\delta \psi^2 (\mathrm{x})$, defined on every patch, only the super-horizon modes  should be taken into account.  This qualitative discussion verifies that for each patch and for times well after the critical point  one can use the approximation
\ba
\delta \psi^2(\mathrm{x},n)  \simeq \delta \psi^2(\mathrm{x},0) ~ e^{2 \kappa_\psi n^2}, \qquad n \gtrsim \frac{1}{\sqrt{2 \kappa_\psi}}.
\ea
The above arguments support our discussions at the main text about the main $n$-dependence of $\delta \psi^2$ for every patch.

\section{Careful treatments of the waterfall induced curvature perturbations}
\label{curvature-rigorous}

Here we study in details the integral in Eq. (\ref{corr-2}) to find the induced curvature perturbations from the waterfall field.

From Eq. (\ref{dpsibetter}) for the mode function $\delta\psi_p$, we have
\ba
\vert \delta\psi_{|\bm{k}-\bm{q}|}(0)\vert^2
= \frac{H^2}{2k_c^3}
 \exp \left[-3n_{|\bm{k}-\bm{q}|} 
-2\kappa_\psi n_{|\bm{k}-\bm{q}|}^2\right]\,,
\ea
where
\ba
n_{|\bm{k}-\bm{q}|}
= \frac{1}{2}\ln\left[\frac{k^2 +q^2 -2qk \cos \theta}{k_c^2}\right]\,,
\ea
and $\kappa_\psi = \beta r /3$.

The momentum integral contain an angular part plus radial part.
Taking account of the azimuthal symmetry, we can write
\ba
\int d^3 q = \int 2 \pi q^2 d q \, \int d (\cos \theta)
\ea
where $\cos\theta=\bm{k}\cdot\bm{q}/kq$.
Changing the integration variable from $\cos\theta$
to $n_{\vert \bm{k}-\bm{q}\vert}$, the angular integral becomes
\ba
\int d (\cos \theta) 
\vert \delta \psi_{\vert \bm{k}-\bm{q}\vert }(0) \vert^2 
= \frac{H^2}{2k_cqk} \, \int _{n_{|k-q|}}^{n_{k+q}}
dn \, \exp \left[ -n-2\kappa_\psi  n^2\right]\,.
\ea
The result of the above integral is in the form of the error functions,
\ba
\label{ang-int}
\int d (\cos \theta) 
\vert \delta \psi_{\vert \bm{k}-\bm{q}\vert }(0) \vert^2
 = \frac{H^2}{2k_cqk} \sqrt{\frac{\pi}{2\kappa_\psi}} 
e^{1/(8\kappa_\psi)}
\left[ \mathrm{Erf} 
\left( \frac{1+4 \kappa_\psi n_{k+q}}{2 
\sqrt{2\kappa_\psi}}\right) -\mathrm{Erf} 
\left( \frac{1+4 \kappa_\psi n_{\vert k-q \vert }}
{2 \sqrt{2\kappa_\psi}}\right) \right]\,,
\ea
where 
\ba
n_{k+q}  &=& \ln \left(\frac{k+q}{k_c} \right) \,,
\\
\label{nk-p}
n_{\vert k-q \vert}& =& \ln \left(\frac{\vert k-q \vert }{k_c} \right)\,,
\ea
and the error function is defined by
\begin{eqnarray}
\mathrm{Erf}(z)=\frac{2}{\sqrt{\pi}}\int_0^z e^{-t^2}dt\,.
\end{eqnarray}

Plugging Eq.~(\ref{ang-int}) into the total momentum integral one has
\ba
\label{tot-int}
\nonumber
&& \big \langle \left( \delta \psi^2\right)_{\bm{k}}
\,\left( \delta \psi^2\right)_{\bm{k}'}\big \rangle  =
\delta^3(\bmk + \bm{k}')\times 
\\
&&2\pi H^2 \sqrt{\frac{\pi}{2\kappa_\psi}} 
e^{1/(8\kappa_\psi)}\int \frac{ q d q}{k_ck }
\vert \delta \psi_q(0)\vert^2 \, \left[ \mathrm{Erf} 
\left( \frac{1+4\kappa_\psi n_{k+q}}
{2 \sqrt{2\kappa_\psi}}\right)
 -\mathrm{Erf} \left( \frac{1+4 \kappa_\psi n_{\vert k-q \vert }}
{2 \sqrt{2\kappa_\psi}}\right) \right].
\ea
The error function, $\mathrm{Erf}(z)$ 
varies from $-1$ for $z\to-\infty$ to $+1$ for $z\to\infty$.
This transition is sharp and take place near $z=0$. 
One can see that the overall behavior of this function is similar 
to $\tanh (z)$. For $k$ not extremely small, the argument of the 
first error function in the square brackets is always much greater
than unity, hence this term is almost $1$. 
The second term depends on the difference between the amplitudes of 
two momenta, $\vert k -q \vert$, and can change from $-1$ for $q=k$
 to $1$ for $q$ completely different from $k$. There is two criteria, first, a very narrow region in which the argument of the second error function become small or negative, in this region the difference between two error functions is considerable but the domain of the integration is very small. Second possibility is when the amplitude of the integrand is not too large but the domain of integration is large. This point is more clarified in the following
 
We showed in the draft that the contribution of the modes in the first criteria is suppressed by a factor $e^{-1/8 \kappa_\psi}$ as follows on the first contribution to the integral. 
\ba
\Big \langle \left( \delta \psi^2\right)_{\bm{k}} 
\,\left( \delta \psi^2\right)_{\bm{k}'}  \Big \rangle_{1}
  &=&2 (2 \pi)^3 \delta^3(\bmk+\bm{k}') \frac{H^2}{4\pi^2}
 \sqrt{\frac{\pi}{2\kappa_\psi}} e^{-1/(8\kappa_\psi)} \,
 \vert \delta \psi_{\bmk}(0) \vert^2 
\ea
It's clear from the steps of the calculation that this suppression is due to narrow width of the integration domain.
\\
It's time to calculate the second contribution. This contribution comes from the region in which the arguments of two error functions in the big bracket are not too far apart and hence the difference between the error functions is small. As we pointed out before, the argument of the first error function is large and it's amplitude of  is approximately equal to unity. outside the first regime discussed above the amplitude of the second error function is also large so we can use the following estimate for the behavior of the error functions with large argument
\ba
\label{erf-app}
\left( 1 - \mathrm{Erf} z\right) \sim \dfrac{e^{-z^2}}{\sqrt{\pi}z}
\ea
Let us first clarify a tiny point. We defined before the first contribution to the integral comes from the region in which
\ba
\frac{1+4 \kappa_\psi n_{\vert k-q \vert }}
{2 \sqrt{2\kappa_\psi}} < z_{bw} \sim 1
\ea
and here we are going to concentrate on the region in which
\ba
\frac{1+4 \kappa_\psi n_{\vert k-q \vert }}
{2 \sqrt{2\kappa_\psi}} > 1
\ea
one can say as $z_{bw} \simeq 1/4 $ you may miss some of contribution to the integral but we claim that this error is not considerable. By using the Eq. (\ref{erf-app}) one has
\ba
 \mathrm{Erf} \left( \frac{1+4\kappa_\psi n_{k+q}}
{2 \sqrt{2\kappa_\psi}}\right)  -\mathrm{Erf} \left( \frac{1+4 \kappa_\psi n_{\vert k-q \vert }}
{2 \sqrt{2\kappa_\psi}}\right) \simeq 2 \sqrt{\dfrac{2 \kappa_\psi}{\pi}} \dfrac{e^{-1/8 \kappa_\psi}}{1+ 4 \kappa_\psi n_{|k-q|}} \exp \left(-n_{|k-q|} -2 \kappa_\psi n_{|k-q|}^2 \right) \nonumber
\ea
One can also more simplify the above equation by ignoring the term $+ 4 \kappa_\psi n_{|k-q|}$ in the denominator of the fist fraction. This approximation can be verified by noting that the total term has as $\exp \left(2 \kappa_\psi n_{|k-q|}^2 \right)$ dependence.

By using the above approximation one has for the second contribution to the integral of the two point correlation function
\ba
\big \langle \left( \delta \psi^2\right)_{\bm{k}}
\,\left( \delta \psi^2\right)_{\bm{k}'}\big \rangle_{2}  =
\delta^3(\bmk + \bm{k}')\times 4\pi H^2 \int \frac{ q d q}{k_ck }
\vert \delta \psi_q(0)\vert^2 \, \exp \left(-n_{|k-q|} -2 \kappa_\psi n_{|k-q|}^2 \right).
\ea
By noting that $n_{|k-q| }= \ln \dfrac{|k-q|}{k_c}$, one has
\ba
d q = \pm k_c e^{n_{|k-q|}} d n_{|k-q|}
\ea
respectively for $q>k $ and $q< k $. Again we claim that as this combination of the error functions are highly peaked around the $q=k$ we take the explicit $q$-dependences out of the integral and identify $q$ in them by $k$. By this trick and noting that the remaining integrand is even one has
\ba
\big \langle \left( \delta \psi^2\right)_{\bm{k}}
\,\left( \delta \psi^2\right)_{\bm{k}'}\big \rangle_{2}  \simeq
\delta^3(\bmk + \bm{k}')\times 8\pi H^2 \vert \delta \psi_k(0)\vert^2 \int  \md n_{|k-q|}
\exp \left(-2 \kappa_\psi n_{|k-q|}^2 \right).
\ea
By performing the integration finally we have
\ba
\big \langle \left( \delta \psi^2\right)_{\bm{k}}
\,\left( \delta \psi^2\right)_{\bm{k}'}\big \rangle_{2}  \simeq
(2 \pi)^3 \delta^3(\bmk + \bm{k}')\times\dfrac{1}{2} \langle \delta \psi ^2(0) \rangle  \vert \delta \psi_k(0)\vert^2 
\ea
This relation is in complete agreement with our previous estimate for the two point correlation function, { Eq.} (\ref{tpcf})


\section{Curvature perturbations for  $k>k_c$ with $\phi_c^2 > 12M_p^2, \kappa_\psi  \gg 1$}

\label{app:deln-k>kc}

In this appendix we are going to study the contribution of the quantum
 fluctuations with $k>k_c$ to the final curvature perturbations for 
the case $\phi_c^2 > 12M_p^2$, using the $\delta N$ formalism. 
For this purpose, we trace back the number of $e$-folds, ${\cal N}$,
from the end of inflation  until time of horizon
 crossing of some specific mode during transition $n$. 
To avoid confusion, in this appendix we denote the number of $e$-folds 
counted backward from end of inflation by $\calN$. Namely,
$\calN=n-n_f$ where $n_f$ is the number of $e$-folds from
the critical point $\phi=\phi_c$ until the end of inflation.
Our strategy is to express ${\cal N}$ in terms of the fields
$\phi(n)$ and $\psi^2(n)$ (smoothed on every Hubble patch).

As discussed in sections~\ref{sec:model} and \ref{sec:background},
in the models with $\phi_c^2 > 12M_p^2$,
 the transition is followed by a chaotic period of inflation. 
So the end of inflation is determined by the value of $\phi$ alone as
\begin{eqnarray}
\phi=\phi_f\approx \sqrt{2/3}\,M_P\,.
\end{eqnarray}
From this up to the end of the waterfall transition
${\cal N}$ is given by
\begin{eqnarray}
{\cal N}=\frac{\phi^2-\phi_f^2}{4M_P^2}
\approx\frac{\phi^2}{4M_P^2}\,;\quad {\cal N}\leq {\cal N}_p\,,
\label{afterNp}
\end{eqnarray}
where ${\cal N}_p$ is the value of $\calN$
at the end of waterfall transition.
As shown in Eq.~(\ref{psi-min-phi}) the value of $\psi^2$ 
at $\calN=\calN_p$ is 
\begin{eqnarray}
\psi^2(n_p)=\psi_{min}^2(n_p)=\frac{M^2-g^2\phi(\calN_p)^2}{\lambda}
=\frac{M^2-4g^2 {\cal N}_pM_P^2}{\lambda}\,.
\end{eqnarray}
Here ${\cal N}_p$ is given in terms of the number of $e$-folds 
from the critical epoch $\phi=\phi_c$ to the end of waterfall transition,
which is denoted by $n_p$, as
\begin{eqnarray}
{\cal N}_p(n_p)=\frac{\phi_p^2}{4M_P^2}
=\frac{\phi_c^2e^{-2rn_p}}{4M_P^2}\,.
\label{Np}
\end{eqnarray}
Hence $\psi^2(\calN_p)$ is a function of $n_p$
through its dependence on $\phi(\calN_p)$,
\begin{eqnarray}
\psi^2(\calN_p)
=\psi_{min}^2(n_p)
=\frac{M^2}{\lambda}-4\frac{g^2M_P^2}{\lambda} {\cal N}_p(n_p)\,.
\label{psip2}
\end{eqnarray}

Now we trace back the evolution to earlier times before the end of 
transition, ${\cal N}>{\cal N}_p$. For this stage, instead
of ${\cal N}$, it is more convenient to use $n$ which is the 
number of $e$-folds from the critical point counted 
{\it forward} in time, ie, $n=n_p+{\cal N}_p- {\cal N}$. 
Then $\psi^2(n)$ is given by 
\begin{eqnarray}
\psi^2(n)=\exp2[f(n)-f(n_p)]\psi_{min}^2(n_p)\,,
\label{psi2}
\end{eqnarray}
where $f(n)$ a function that describes the tachyonic growth
of the waterfall field, as given by Eq.~(\ref{fnform}).
During this era, as for $\phi$, we have
\begin{eqnarray}
\phi(n)=\phi_ce^{-rn}\,.
\label{phi}
\end{eqnarray}
Here we note that $n$ depends on $n_p$ and ${\cal N}$ in
 a rather non-trivial way,
\begin{eqnarray}
n(n_p,{\cal N})=n_p+ {\cal N}_p(n_p)- {\cal N}\,.
\label{smalln}
\end{eqnarray}
In particular, we have
\begin{eqnarray}
\frac{\partial n}{\partial n_p}=1+\frac{d {\cal N}_p}{dn_p}
=1-2r {\cal N}_p\,,
\end{eqnarray}
where we have used the fact that
\begin{eqnarray}
\frac{d {\cal N}_p}{dn_p}=-2r {\cal N}_p\,,
\end{eqnarray}
which follows from Eq.~(\ref{Np}).

Keeping in mind the above dependence of $n$ on $n_p$ and ${\cal N}$,
let us take the variation of (\ref{psi2}) and (\ref{phi}). We obtain
\begin{eqnarray}
\delta\psi^2(n)
&=&-2\psi^2(n)f'(n)\delta {\cal N}
\label{dpsi2}\\
&&\quad
+\psi^2(n)\left(2\left[f'(n)\frac{\partial n}{\partial n_p}-f'(n_p)\right]
+\frac{d\ln\psi_{min}^2(n_p)}{dn_p}\right)\delta n_p\,.
\cr
\cr
\delta\phi(n)
&=&-r\phi(n)\delta n
=r\phi(n)\delta {\cal N} -r\phi(n)\frac{\partial n}{\partial n_p}\delta n_p\,.
\label{dphi}
\end{eqnarray}
By using Eq.~(\ref{psip2}), one finds
\begin{eqnarray}
\left|\frac{d\ln\psi_{min}^2(n_p)}{dn_p}\right| 
\simeq 8 g^2 \dfrac{M_P^2}{M^2} \, r\calN_p \lesssim 2 r \ll 1\,,
\end{eqnarray}
where we have used Eq.~(\ref{afterNp}) and the fact that 
$\phi_p \lesssim\phi_c$ in the last step.

To find the variations of the number of $e$-folds, one
solves Eqs.~(\ref{dpsi2}) and (\ref{dphi}) for $\delta {\cal N}$
to obtain,
\begin{eqnarray}
\delta {\cal N}=\frac{\delta\psi^2(n)}{\psi^2(n)}
\dfrac{\partial n}{\partial n_p}
\frac{1}{-2f'(n_p)}
+\frac{\delta\phi(n)}{r\phi(n)}
\left[1+
\frac{2f'(n)}
{-2f'(n_p)}\frac{\partial n}{\partial n_p}
\right]\,,
\end{eqnarray}
in which we have ignored the term $|d\ln\psi_p^2(n_p)/dn_p| \sim r \ll1$.
Note that as a result of the vacuum domination condition, 
we also have $\left|d {\cal N}_p/dn_p\right|=2r {\cal N}_p\ll1$.
Then the above reduces to
\begin{eqnarray}
\delta \calN=-\frac{1}{2f'(n_p)}\frac{\delta\psi^2(n)}{\psi^2(n)}
+\left[1-\frac{f'(n)}{f'(n_p)}\right]\frac{\delta\phi(n)}{r\phi(n)}\,.
\end{eqnarray}
As clear from this result, the curvature perturbation spectrum is
smoothly taken over from $\delta\phi$ to $\delta\psi^2$ as 
$n$ increases from 0 to $n_p$.


\end{document}